\documentclass[aps,pre,twocolumn,amssymb,amsfonts,floatfix,superscriptaddress,longbibliography,notitlepage,footinbib]{revtex4-1}
\usepackage[utf8]{inputenc}
\pdfoutput=1
\usepackage{color}
\usepackage{float}
\usepackage{xcolor}
\usepackage{mathtools,amsmath,amsthm,amsfonts,amssymb}
\usepackage{commath}
\usepackage{physics}
\usepackage{bm}
\usepackage[pdftex]{graphicx}
\usepackage[colorlinks=false]{hyperref}
\usepackage{enumitem}
\usepackage{soul}

\begin{document}

\title{Superradiant Phase is a Finite Size Effect in Two-photon Processes}

\author{Fabrizio Ram\'irez}
\affiliation{Instituto de Ciencias Nucleares, Universidad Nacional Aut\'onoma de M\'exico, Apdo. Postal 70-543, C.P. 04510 Mexico City, Mexico}
\author{David Villase\~nor}
\email{d.v.pcf.cu@gmail.com}
\affiliation{CAMTP - Center for Applied Mathematics and Theoretical Physics, University of Maribor, Mladinska 3, SI-2000 Maribor, Slovenia}
\author{Nahum V\'azquez}
\author{Jorge G. Hirsch}
\email{hirsch@nucleares.unam.mx}
\affiliation{Instituto de Ciencias Nucleares, Universidad Nacional Aut\'onoma de M\'exico, Apdo. Postal 70-543, C.P. 04510 Mexico City, Mexico}

\begin{abstract}
Two-photon light–matter interactions exhibit distinctive features such as spectral collapse. The two-photon Dicke model has been reported to exhibit a superradiant phase which could be useful in quantum applications. Here we show that this superradiant phase is not a genuine thermodynamic phase but a finite-size effect. Combining analytical and numerical analyses, we demonstrate that the superradiant region shrinks with increasing system size and disappears in the thermodynamic limit, while spectral collapse remains. Our results clarify the nature of superradiant conditions in two-photon systems and constrain its realization in quantum platforms.
\end{abstract}

\maketitle

Understanding the interaction between light and matter is a central problem in quantum optics~\cite{Scully2008Book} and underpins a wide range of quantum technologies, including quantum metrology~\cite{Taylor2016,Degen2017,Fiderer2018,Crawford2021,Montenegro2025}, simulation~\cite{Lloyd1996,Georgescu2014}, and communication~\cite{Zeilinger1999,Gisin2007,Swingle2016,Landsman2019}. Cavity and circuit quantum electrodynamics provide controlled settings in which collective atom–field effects emerge~\cite{Zeilinger1999,Dowling2003}, often leading to enhanced coherence and cooperative phenomena that have no single-particle counterpart.

Among the paradigmatic models describing collective light–matter interactions is the Dicke model~\cite{Dicke1954,Kirton2019,Roses2020,Larson2021Book,Villasenor2024ARXIV}, which captures the coupling between an ensemble of two-level systems (qubits) and a single bosonic mode. The Dicke model exhibits a well-established superradiant phase transition in the thermodynamic limit~\cite{Emary2003,Emary2003PRL,Lambert2004,Brandes2005,Brandes2013}, characterized by macroscopic occupation of the field mode and collective atomic excitation. This transition has been extensively studied both theoretically~\cite{Rehler1971,Bonifacio1971,MacGillivray1976,Gross1982,Sutherland2017,Masson2020,Masson2022,Sierra2022,Pineiro2022,Cardenas2023,Masson2024,Rosario2025} and experimentally~\cite{Skribanowitz1973,Raimond1982,Inouye1999,Slama2007,Scheibner2007} and forms a cornerstone of the understanding of collective quantum behavior.

Recent interest has shifted toward multiphoton light–matter interactions~\cite{Lo1998,Braak2026}, motivated by advances in experimental platforms such as trapped ions, superconducting circuits, Rydberg atoms, quantum dots, and quantum batteries~\cite{Loy1978,Schlemmer1980,Nikolaus1981,Brune1987,Brune1987PRL,Brune1988,Gauthier1992,Stufler2006,DelValle2010,Felicetti2018,Crescente2020,Wang2024}. In particular, two-photon processes, where atomic excitations are accompanied by the simultaneous creation or annihilation of two bosonic excitations, lead to qualitatively new physics~\cite{Agarwal1985,Alsing1987,Puri1988,Toor1992,Lo1998,Puri1988PRA,Braak2026}. The two-photon Rabi and Dicke models display rich spectral properties, including spectral collapse~\cite{Felicetti2015,Duan2016,Rico2020,Lo2021}, in which discrete energy levels merge into a continuous band at a critical coupling strength. This phenomenon has no analogue in single-photon models.

The two-photon Dicke model~\cite{Gerry1989,Felicetti2015,Peng2017,Wang2019TPD,Larson2021Book,Lo2021,Banerjee2022,Ramirez2025} has attracted special attention because it has been reported to exhibit a superradiant phase~\cite{Garbe2017,Chen2018} based on analytical descriptions using Holstein–Primakoff and squeezed-state mean-field approaches. This superradiant phase has been found even in the presence of dissipation~\cite{Garbe2020,Shah2025}. It has motivated proposals for exploiting two-photon interactions in quantum technologies expected to operate under superradiant conditions~\cite{Crescente2020,Wang2024,Zhai2025}.
However, the nature of this superradiant phase and its stability in the thermodynamic limit have remained insufficiently scrutinized.

In this Letter, we show that the superradiant phase of the two-photon Dicke model is not a genuine thermodynamic phase but rather a finite-size effect. By combining analytical mean-field results with numerical simulations, we demonstrate that the region associated with superradiant behavior systematically shrinks as the system size increases and vanishes entirely in the thermodynamic limit. Employing a mean-field description based on squeezed vacuum states, we show that the solutions of the superradiant phase are unphysical in the thermodynamic limit.

Our results clarify the relationship between superradiance and spectral collapse in two-photon light–matter systems. We find that spectral collapse survives the thermodynamic limit and admits a clear interpretation in bosonic phase space, whereas superradiance does not. This distinction resolves an apparent inconsistency in previous analyses and imposes fundamental constraints on the realization of superradiant behavior in two-photon platforms.

\begin{figure*}[ht]
    \centering
    \includegraphics[width=1\textwidth]{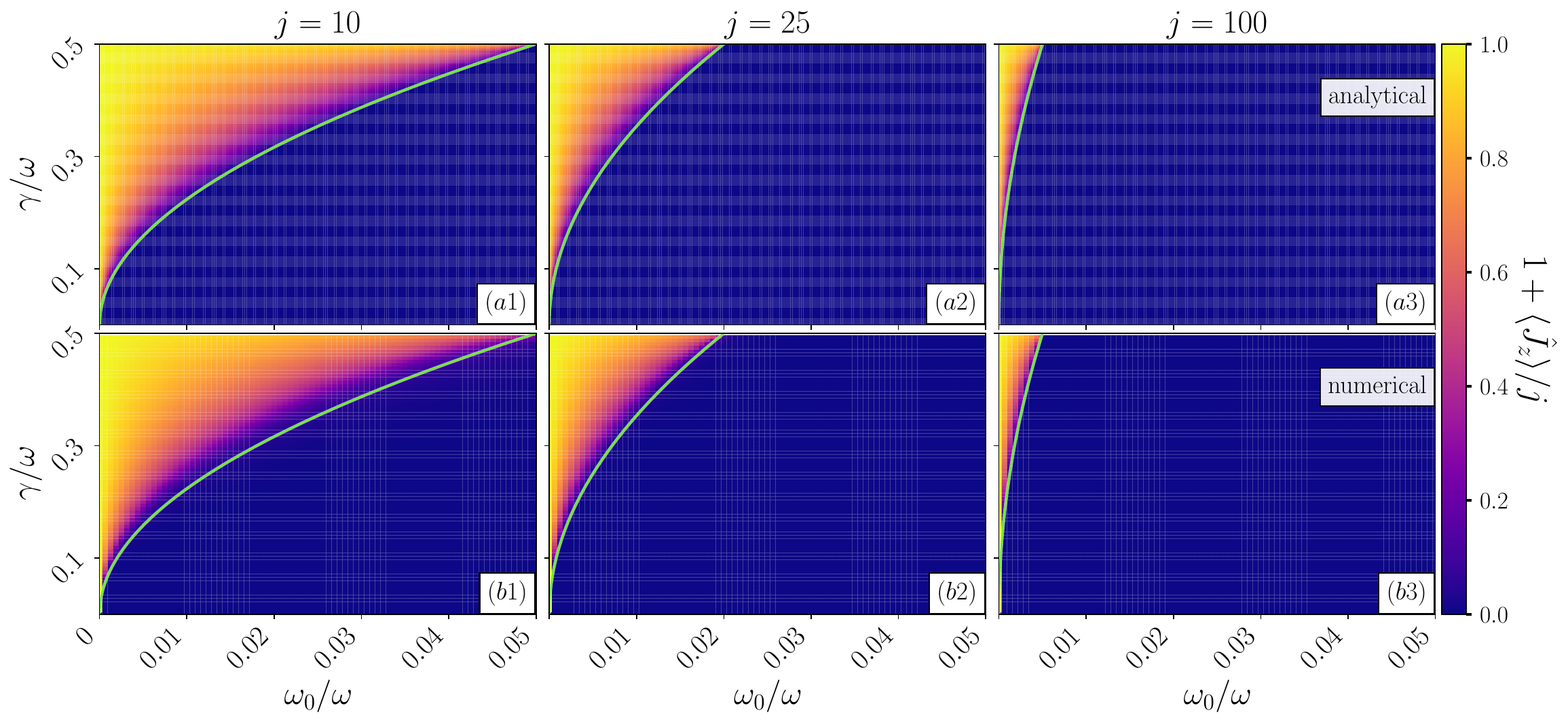}
    \caption{(a1)-(a3) Phase diagrams of the normalized atomic excitation $1 + \langle \hat{J}_{z} \rangle/j$ [Eq.~\eqref{eq:JzVariable}] as a function of the scaled atomic frequency $\omega_{0}/\omega$ and the scaled coupling strength $\gamma/\omega$. The color scale represents the numerical value of the normalized atomic excitation. The green solid line represents the scaled critical coupling $\gamma_{\text{c}}/\omega=\sqrt{\omega_{0}j/(2\omega)}$. (b1)-(b3) Numerical implementations of panels (a1)-(a3). Each column identifies a different system size: (a1),(b1) $j=10$, (a2),(b2) $j=25$, and (a3),(b3) $j=100$.
    }
    \label{fig:JzFadingPhase}
\end{figure*}

{\it Two-photon Dicke model---}
The two-photon Dicke model describes the collective interaction between
$\mathcal{N}$ two-level systems and a single-mode bosonic field of frequency 
$\omega$, where atomic excitations are accompanied by the simultaneous creation or annihilation of two bosonic quanta. Setting $\hbar=1$, the Hamiltonian reads as
\begin{equation}
    \label{eq:DickeHamiltonian}
    \hat{H}_{\text{D}} = \omega \hat{a}^{\dagger} \hat{a} + \omega_{0} \hat{J}_{z} + \frac{\gamma}{\mathcal{N}} \left( \hat{a}^{\dagger 2} + \hat{a}^{2} \right)\left( \hat{J}_{+} + \hat{J}_{-} \right) ,
\end{equation}
where $\hat{a}^\dagger$ and $\hat{a}$ are the creation and annihilation operators satisfying $[\hat{a}, \hat{a}^{\dagger}] = \hat{\mathbb{I}}$, respectively. The collective atomic operators $\hat{J}_{z}$ and $\hat{J}_{\pm}=\hat{J}_{x}\pm i\hat{J}_{y}$ obey the standard angular-momentum commutation relations $[\hat{J}_{z},\hat{J}_{\pm}]=\pm\hat{J}_{\pm}$ and $[\hat{J}_{+},\hat{J}_{-}]=2\hat{J}_{z}$ and are defined as $\hat{J}_{x,y,z} = (1/2)\sum_{k = 1}^{\mathcal{N}}\hat{\sigma}_{x,y,z}^{k}$. In the above Hamiltonian, $\omega_{0}$ denotes the atomic transition frequency, and $\gamma$ is the atom–field coupling strength. The coupling is bounded by the critical value $\gamma_{\text{sc}}=\omega/2$, at which spectral collapse occurs~\cite{Felicetti2015,Duan2016,Rico2020}.

The Hamiltonian $\hat{H}_{\text{D}}$ possesses a discrete parity symmetry, $[\hat{H}_{\text{D}},\hat{\Pi}]=0$, where the parity operator is defined as 
 $\hat{\Pi} = \text{exp}(i\pi\hat{\Lambda})$, with $\hat{\Lambda}=\hat{a}^{\dagger}\hat{a}/2 + \hat{J}_{z} + j\hat{\mathbb{I}}$. This symmetry partitions the Hilbert space into four invariant subspaces labeled by the eigenvalues $p=\pm 1,\pm i$ of $\hat{\Pi}$, corresponding to the simultaneous transformations $\hat{a} \to i \hat{a}$ and $\hat{J}_x \to -\hat{J}_x$.

In addition, the total collective spin $\hat{\mathbf{J}}^{2}=\hat{J}_{x}^{2}+\hat{J}_{y}^{2}+\hat{J}_{z}^{2}$ is conserved, $[\hat{H}_{\text{D}},\hat{\mathbf{J}}^{2}]=0$, allowing the Hilbert space to be decomposed into sectors characterized by the eigenvalue $j(j+1)$. Throughout this work, we focus on the fully symmetric subspace with $j=\mathcal{N}/2$, which governs the collective behavior of the model and is relevant for the thermodynamic limit.

\begin{figure*}[ht]
    \centering
    \includegraphics[width=1\textwidth]{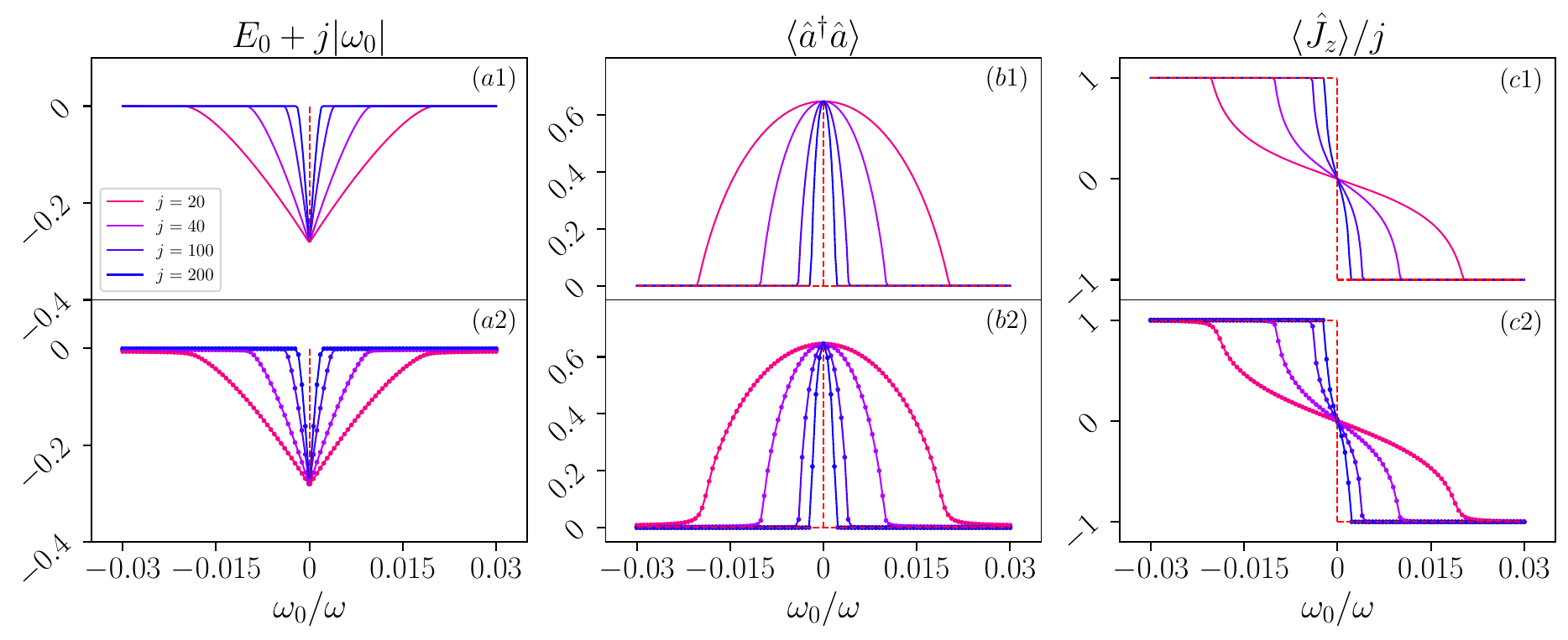}
    \caption{Projections of (a1) displaced ground-state energy $E_{0} + j|\omega_{0}|$ [Eq.~\eqref{eq:GroundStateEnergy}], (b1) number of photons $\langle \hat{a}^{\dagger}\hat{a} \rangle$ [Eq.~\eqref{eq:PhotonNumber}], and (c1) atomic operator $\langle \hat{J}_{z} \rangle/j$ [Eq.~\eqref{eq:JzVariable}], as a function of the scaled atomic frequency $\omega_{0}/\omega$. Each color identifies a different system size $j=20,40,100,200$. The red dashed line represents the thermodynamic limit $j\to\infty$. (a2)-(c2) Numerical implementations of panels (a1)-(c1). The coupling strength is $\gamma/\omega=0.45$.
    }
    \label{fig:Slices}
\end{figure*}

{\it Ground state, classical limit, and superradiant phase---}
The ground state of the two-photon Dicke model can be obtained at the mean-field level by applying a Holstein–Primakoff transformation to the collective spin operators, followed by a Bogoliubov transformation of the bosonic mode~\cite{Garbe2017}. This procedure yields a squeezed-vacuum ground state of the form $|E_{0}\rangle = |r_{b}\rangle = e^{(r^{\ast}_{b}\hat{a}^{2}-r_{b}\hat{a}^{\dagger2})/2}|0\rangle$, where $r_{b}\in\mathbb{R}$ is a real squeezing parameter and $|0\rangle$ 
denotes the bosonic vacuum. Minimization of the ground-state energy $E_{0}$  identifies two regimes separated by the critical coupling $\gamma_{\text{c}}=\sqrt{\omega\omega_{0}j/2}$. For $\gamma<\gamma_{\text{c}}$ the system is in a normal phase (NP), while for $\gamma>\gamma_{\text{c}}$ a superradiant phase (SP) emerges. Details of this derivation are given in Ref.~\cite{Garbe2017}.

The previous mean-field description does not allow for analyzing the scaling of observables with the number of atoms, nor does it yield the thermodynamic limit. Thus, we introduce a similar mean-field approximation based on the tensor product $|\mathbf{x}\rangle \equiv |\xi\rangle\otimes|\beta\rangle$, where $|\xi\rangle=e^{(\xi^{\ast}\hat{a}^{2}-\xi\hat{a}^{\dagger2})/2}|0\rangle$ is a squeezed vacuum state and $|\beta\rangle = (1+|\beta|^{2})^{-j}e^{\beta\hat{J}_{+}}|j,-j\rangle$ is a Bloch coherent state for the collective spin. The squeezing parameter is a complex number $\xi=re^{i\theta}\in\mathbb{C}$ with magnitude $r$ and phase $\theta$. We introduce the canonical variables $\mathbf{x}=(q,p;Q,P)$ describing the bosonic and atomic degrees of freedom. The atomic variables $(Q,P)$ are related to the Bloch parameter $\beta = (Q+iP)/\sqrt{4-Q^{2}-P^{2}}$, while the bosonic variables $(q,p)$ are
\begin{equation}
    \label{eq:BosonicVariables}
    (q,p) = \sqrt{\frac{2}{j}}\sinh(r)[\cos(\theta),\sin(\theta)] .
\end{equation}

The scaling factor $\sqrt{2/j}$ in Eq.~\eqref{eq:BosonicVariables} allows us to define a classical Hamiltonian,
\begin{align}
    \label{eq:ClassicalDickeVacuum}
    h_{\text{D}}(\mathbf{x}) = & \frac{1}{j}\langle \mathbf{x}|\hat{H}_{\text{D}}|\mathbf{x} \rangle \\
    = &  \Omega(\mathbf{x}) - 2\gamma qQ\sqrt{\left(\frac{1}{2j}+\frac{q^{2}+p^{2}}{4}\right)\left(1-\frac{Q^{2}+P^{2}}{4}\right)} , \nonumber
\end{align}
with
\begin{equation}
    \label{eq:IntegrableHamiltonian}
    \Omega(\mathbf{x}) = \frac{\omega}{2}\left(q^{2}+p^{2}\right) + \frac{\omega_{0}}{2}\left(Q^{2}+P^{2}\right) - \omega_{0} ,
\end{equation}
where the energy is scaled with the number of atoms $j=\mathcal{N}/2$, namely $h_{\text{D}}(\mathbf{x})=E/j$, and its thermodynamic limit provides an intensive classical energy.

The critical points from Hamilton’s equations of motion of the Hamiltonian $h_{\text{D}}(\mathbf{x})$ are $\mathbf{x}_{m}$~\cite{footSM}. One of these extrema $\mathbf{x}_{0}\in\mathbb{R}$ is related to the normal phase, while the remaining extrema $\mathbf{x}_{\pm}\in\mathbb{R}$ are related to the superradiant phase. Using these extrema, we can obtain analytical expressions~\cite{footSM} for the ground-state energy,
\begin{align}
\label{eq:GroundStateEnergy}
    E_{0} = & \langle \mathbf{x}_{m}| \hat{H}_{\text{D}} |\mathbf{x}_{m} \rangle \\
    = & -\left\{\begin{array}{ll}
       j|\omega_{0}| & \text{for (NP)} \\
        \frac{j\omega}{2}\left(\frac{1}{j} - \sqrt{\left(1-\frac{4\gamma^{2}}{\omega^{2}}\right)\left(\frac{1}{j^{2}}-\frac{\omega_{0}^{2}}{\gamma^{2}}\right) }\right) & \text{for (SP)}
    \end{array}\right. \nonumber ,
\end{align}
the number of photons
\begin{equation}
\label{eq:PhotonNumber}
    \langle \mathbf{x}_{m}| \hat{a}^{\dagger}\hat{a} |\mathbf{x}_{m} \rangle = \left\{\begin{array}{ll}
        0 & \text{for (NP)} \\
        \frac{j}{2}\left(\sqrt{\frac{\frac{1}{j^{2}}-\frac{\omega_{0}^{2}}{\gamma^{2}}}{1-\frac{4\gamma^{2}}{\omega^{2}}}} - \frac{1}{j}\right) & \text{for (SP)}
    \end{array}\right. ,
\end{equation} 
and the atomic inversion
\begin{equation}
    \label{eq:JzVariable}
    \frac{\langle \mathbf{x}_{m}| \hat{J}_{z} |\mathbf{x}_{m} \rangle}{j} = -\left\{\begin{array}{ll}
        \text{sgn}(\omega_{0}) & \text{for (NP)} \\
        \frac{\omega_{0}\omega}{2\gamma^{2}}\sqrt{\frac{1-\frac{4\gamma^{2}}{\omega^{2}}}{\frac{1}{j^{2}}-\frac{\omega_{0}^{2}}{\gamma^{2}}}} & 
        \text{for (SP)}
    \end{array}\right. .
\end{equation}

\begin{figure*}[ht]
    \centering
    \includegraphics[width=0.9\textwidth]{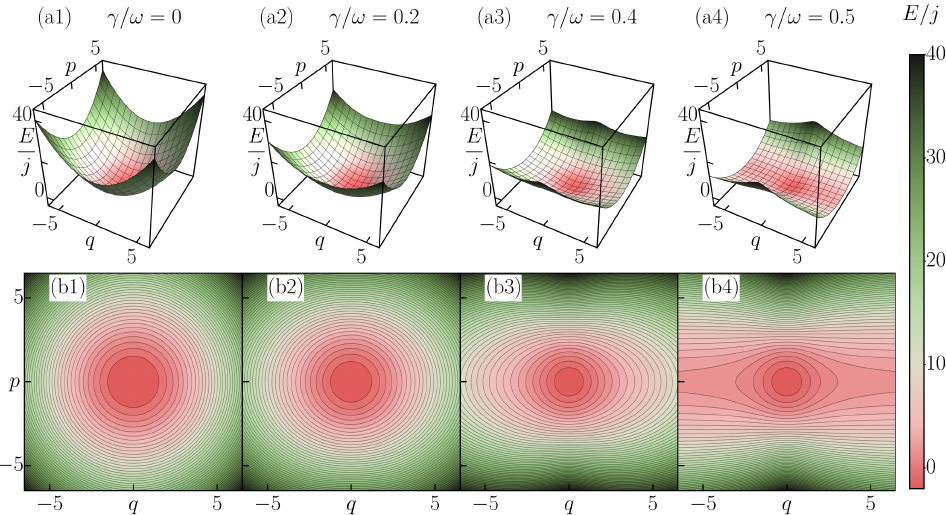}
    \caption{(a1)-(a4) Energy surface $h_{\text{D}}^{\infty}(q,p)$ [Eq.~\eqref{eq:E_surface_bosonic}] as a function of the bosonic coordinates $(q,p)$ for increasing values of the scaled coupling strength $\gamma/\omega=0.1,0.2,0.4,0.5$. (b1)-(b4) Orthogonal projections of panels (a1)-(a4) over the bosonic plane $q-p$. The color scale shows values of the scaled classical energy $E/j$ that begins at the scaled ground-state energy $E_{0}/j=-\omega_{0}$. System parameters: $\omega=1$ and $\omega_{0}=2\omega$.}
    \label{fig:SpectralCollapse}
\end{figure*}

{\it Thermodynamic limit and disappearance of the superradiant phase---}
The thermodynamic limit of the two-photon Dicke model is defined by $\mathcal{N}\to\infty$ or, equivalently, when $j =\mathcal{N}/2\to\infty$. In this limit, collective atom–field systems are expected to display macroscopic behavior and well-defined phase transitions. The analytical expressions derived in Eqs.~\eqref{eq:GroundStateEnergy}-\eqref{eq:JzVariable} allow us to track explicitly how the ground-state properties evolve with increasing system size.

Figures~\ref{fig:JzFadingPhase}(a1)-\ref{fig:JzFadingPhase}(a3) present phase diagrams of the normalized atomic excitation  $1+\langle\hat{J}_{z}\rangle/j$ [Eq.~\eqref{eq:JzVariable}] as functions of the scaled atomic frequency $\omega_{0}/\omega$ and the scaled coupling strength $\gamma/\omega$ for increasing values of $j$. For finite system sizes, two distinct regions corresponding to the normal and superradiant phases are separated by the critical coupling $\gamma_{\text{c}}=\sqrt{\omega\omega_{0}j/2}$. However, as the system size increases, the superradiant region shrinks progressively. In the limit $j\to\infty$, this region collapses entirely, leaving only the normal phase. The numerical results in Figs.~\ref{fig:JzFadingPhase}(b1)-\ref{fig:JzFadingPhase}(b3), obtained with the ground state from the diagonalization of $\hat{H}_{\text{D}}$ [Eq.~\eqref{eq:DickeHamiltonian}], are in complete agreement with the analytical predictions. We present similar phase diagrams for the number of photons $\langle\hat{a}^{\dagger}\hat{a}\rangle$ [Eq.~\eqref{eq:PhotonNumber}] and the ground-state energy $E_{0}$ [Eq.~\eqref{eq:GroundStateEnergy}] in Supplemental Material~\cite{footSM}. 

The previous behavior is further illustrated in Figs.~\ref{fig:Slices}(a1)-\ref{fig:Slices}(c1), which show projections of the ground-state energy, photon number, and atomic inversion as functions of $\omega_{0}/\omega$ at fixed coupling $\gamma/\omega < 1/2$, respectively. For finite $j$, these observables display the smooth crossover associated with the analytically defined superradiant phase. As $j$ increases, the crossover region narrows and converges to a singular structure in the thermodynamic limit. At $\omega_{0}=0$, the ground-state energy develops a finite discontinuity, the photon number approaches a deltalike peak, and the atomic inversion becomes a step function. These limiting behaviors are incompatible with the existence of a stable superradiant phase. The numerical results in Figs.~\ref{fig:Slices}(a2)-\ref{fig:Slices}(c2) again show full agreement with the analytical results. 

The disappearance of the superradiant phase can be understood directly from the analytical expressions. In Eqs.~\eqref{eq:GroundStateEnergy}-\eqref{eq:JzVariable}, the critical points associated with the superradiant phase require $j^{2}\omega_{0}^{2}<\gamma^{2}$, a condition that cannot be satisfied as $j\to\infty$ at fixed $\gamma$ and $\omega_0$. Consequently, the corresponding mean-field extrema $\mathbf{x}_{\pm}$ become complex and unphysical in the thermodynamic limit, signaling the breakdown of the superradiant-phase solutions. This analysis can also be confirmed by taking the thermodynamic limit in Eqs.~\eqref{eq:GroundStateEnergy}-\eqref{eq:JzVariable}, leading to complex expressions for the three scaled observables~\cite{footSM}.

From the previous evidence, we conclude that the superradiant phase does not survive the thermodynamic limit in the two-photon Dicke model. The phase identified at finite system sizes represents a finite-size crossover rather than a genuine quantum phase transition.

{\it Spectral collapse in bosonic phase space---}
Despite the fact that the superradiant phase does not survive in the thermodynamic limit of the two-photon Dicke model, the classical signature of the spectral collapse can be identified in the same limit. In the limit $j\to\infty$, the term $1/(2j)$ disappears in Eq.~\eqref{eq:ClassicalDickeVacuum} and a new Hamiltonian, which is independent of the number of atoms $j=\mathcal{N}/2$, can be defined as
\begin{align}
    \label{eq:ClassicalDickeVacuumInfinity}
    h_{\text{D}}^{\infty}(\mathbf{x}) = & \lim_{j\to\infty}h_{\text{D}}(\mathbf{x}) \\
    = &  \Omega(\mathbf{x}) - \gamma qQ\sqrt{\left(q^{2}+p^{2}\right)\left(1-\frac{Q^{2}+P^{2}}{4}\right)} , \nonumber
\end{align}
where $\Omega(\mathbf{x})$ is given in Eq.~\eqref{eq:IntegrableHamiltonian}.

The critical points from Hamilton's equations of motion of the Hamiltonian $h_{\text{D}}^{\infty}(\mathbf{x})$ are $\mathbf{x}_{0}\in\mathbb{R}$ (normal phase) and $\mathbf{x}_{\pm}^{\infty}\in\mathbb{C}$ (superradiant phase)~\cite{footSM}, where the only real solution $\mathbf{x}_{0}$ reproduces the ground-state energy of the normal phase, $h_{\text{D}}^{\infty}(\mathbf{x}_{0})=E_{0}/j=-\omega_{0}$. The previous critical points can also be obtained by taking the thermodynamic limit of the critical points that come from the Hamiltonian $h_{\text{D}}(\mathbf{x})$ [Eq.~\eqref{eq:ClassicalDickeVacuum}], namely, $\lim_{j\to\infty}\mathbf{x}_{0}=\mathbf{x}_{0}\in\mathbb{R}$ and $\lim_{j\to\infty}\mathbf{x}_{\pm}=\mathbf{x}_{\pm}^{\infty}\in\mathbb{C}$.

The Hamiltonian $h_{\text{D}}^{\infty}(\mathbf{x})$ allows us to visualize the spectral collapse in the bosonic phase space of the two-photon Dicke model. Eliminating the atomic variables by solving  $\partial_{Q}h_{\text{D}}^{\infty}(\mathbf{x})=\partial_{P}h_{\text{D}}^{\infty}(\mathbf{x})=0$, we obtain the solutions $Q_{m}=Q_{m}(q,p)$ and $P_{m}=0$, leading to an effective energy surface projected onto the bosonic variables
\begin{align}
    \label{eq:E_surface_bosonic}
    h_{\text{D}}^{\infty}(q,p) = & h_{\text{D}}^{\infty}(q,p;Q_{m},0) \\
    = & \frac{\omega}{2}\left(q^{2}+p^{2}\right)-\omega_{0}\sqrt{1+\frac{\gamma^{2}}{\omega_{0}^{2}}q^{2}\left(q^{2}+p^{2}\right)} \, \nonumber .
\end{align}

Figures~\ref{fig:SpectralCollapse}(a1)-\ref{fig:SpectralCollapse}(a4) show the resulting energy surface for increasing values of the scaled coupling strength $\gamma/\omega$ at fixed resonance $\omega_{0}=2\omega$. In Figs.~\ref{fig:SpectralCollapse}(b1)-\ref{fig:SpectralCollapse}(b4), we show the orthogonal projection of the energy surfaces over the bosonic plane $q-p$. For weak coupling, the surface is bounded and displays a single global minimum, characteristic of the normal phase. As $\gamma/\omega$ approaches the collapse value $\gamma_{\text{sc}}/\omega=1/2$, the surface progressively flattens and eventually becomes unbounded, developing extended regions at the ground-state energy across the $p$ axis [Figs.~\ref{fig:SpectralCollapse}(a4) and~\ref{fig:SpectralCollapse}(b4)]. This transition from a bounded to an unbounded energy landscape provides a clear classical signature of spectral collapse in bosonic phase space. 

A complementary description using a mean-field description with coherent states is presented in Supplemental Material~\cite{footSM}. In this case the ground state is always in the normal phase. There are, however, two relevant aspects which are missed in the coherent state description: the presence of the squeezed superradiant phase for finite number of atoms and the squeezed ground state in the limit $\omega_0 \rightarrow 0$.

{\it Conclusion---}
We demonstrated that the superradiant phase of the two-photon Dicke model is not a genuine thermodynamic phase but a finite-size effect. Although a superradiant region emerges at finite system size, it collapses systematically as the number of atoms increases and disappears in the thermodynamic limit. In this limit, the critical points associated with superradiance $\mathbf{x}_{\pm}^{\infty}\in\mathbb{C}$ become unphysical, and the classical Hamiltonian $h_{\text{D}}^{\infty}(\textbf{x})$ admits only the normal phase and correctly captures the onset of spectral collapse.

Our results demonstrate that superradiance and spectral collapse are fundamentally distinct phenomena in two-photon light–matter systems. In particular, spectral collapse survives the thermodynamic limit, whereas superradiance does not. This resolves an apparent contradiction in previous analyses based on squeezed-state mean-field descriptions and places strong constraints on the realization of superradiant behavior in two-photon platforms. Whether dissipation~\cite{Garbe2020,Shah2025} can stabilize a genuine superradiant phase remains an open and nontrivial question.

{\it Acknowledgments---}
We acknowledge the support of the Computation Center - ICN, in particular, Enrique Palacios, Luciano D\'iaz, and Eduardo Murrieta. F.R., N.V., and J.G.H acknowledge partial financial support from the DGAPA-UNAM Projects No. IN109523 and No. IN101526. D.V. acknowledges financial support from the Slovenian Research and Innovation Agency (ARIS) under Grants No. J1-4387 and No. P1-0306.

\bibliography{main}

\begin{thebibliography}{80}%
\makeatletter
\providecommand \@ifxundefined [1]{%
 \@ifx{#1\undefined}
}%
\providecommand \@ifnum [1]{%
 \ifnum #1\expandafter \@firstoftwo
 \else \expandafter \@secondoftwo
 \fi
}%
\providecommand \@ifx [1]{%
 \ifx #1\expandafter \@firstoftwo
 \else \expandafter \@secondoftwo
 \fi
}%
\providecommand \natexlab [1]{#1}%
\providecommand \enquote  [1]{``#1''}%
\providecommand \bibnamefont  [1]{#1}%
\providecommand \bibfnamefont [1]{#1}%
\providecommand \citenamefont [1]{#1}%
\providecommand \href@noop [0]{\@secondoftwo}%
\providecommand \href [0]{\begingroup \@sanitize@url \@href}%
\providecommand \@href[1]{\@@startlink{#1}\@@href}%
\providecommand \@@href[1]{\endgroup#1\@@endlink}%
\providecommand \@sanitize@url [0]{\catcode `\\12\catcode `\$12\catcode `\&12\catcode `\#12\catcode `\^12\catcode `\_12\catcode `\%12\relax}%
\providecommand \@@startlink[1]{}%
\providecommand \@@endlink[0]{}%
\providecommand \url  [0]{\begingroup\@sanitize@url \@url }%
\providecommand \@url [1]{\endgroup\@href {#1}{\urlprefix }}%
\providecommand \urlprefix  [0]{URL }%
\providecommand \Eprint [0]{\href }%
\providecommand \doibase [0]{http://dx.doi.org/}%
\providecommand \selectlanguage [0]{\@gobble}%
\providecommand \bibinfo  [0]{\@secondoftwo}%
\providecommand \bibfield  [0]{\@secondoftwo}%
\providecommand \translation [1]{[#1]}%
\providecommand \BibitemOpen [0]{}%
\providecommand \bibitemStop [0]{}%
\providecommand \bibitemNoStop [0]{.\EOS\space}%
\providecommand \EOS [0]{\spacefactor3000\relax}%
\providecommand \BibitemShut  [1]{\csname bibitem#1\endcsname}%
\let\auto@bib@innerbib\@empty
\bibitem [{\citenamefont {Scully}\ and\ \citenamefont {Zubairy}(1997)}]{Scully2008Book}%
  \BibitemOpen
  \bibfield  {author} {\bibinfo {author} {\bibfnamefont {Marlan~O.}\ \bibnamefont {Scully}}\ and\ \bibinfo {author} {\bibfnamefont {M.~Suhail}\ \bibnamefont {Zubairy}},\ }\href@noop {} {\emph {\bibinfo {title} {Quantum optics}}}\ (\bibinfo  {publisher} {Cambridge University Press},\ \bibinfo {address} {Cambridge},\ \bibinfo {year} {1997})\BibitemShut {NoStop}%
\bibitem [{\citenamefont {Taylor}\ and\ \citenamefont {Bowen}(2016)}]{Taylor2016}%
  \BibitemOpen
  \bibfield  {author} {\bibinfo {author} {\bibfnamefont {Michael~A.}\ \bibnamefont {Taylor}}\ and\ \bibinfo {author} {\bibfnamefont {Warwick~P.}\ \bibnamefont {Bowen}},\ }\bibfield  {title} {\enquote {\bibinfo {title} {Quantum metrology and its application in biology},}\ }\href {\doibase https://doi.org/10.1016/j.physrep.2015.12.002} {\bibfield  {journal} {\bibinfo  {journal} {Phys. Rep.}\ }\textbf {\bibinfo {volume} {615}},\ \bibinfo {pages} {1--59} (\bibinfo {year} {2016})}\BibitemShut {NoStop}%
\bibitem [{\citenamefont {Degen}\ \emph {et~al.}(2017)\citenamefont {Degen}, \citenamefont {Reinhard},\ and\ \citenamefont {Cappellaro}}]{Degen2017}%
  \BibitemOpen
  \bibfield  {author} {\bibinfo {author} {\bibfnamefont {C.~L.}\ \bibnamefont {Degen}}, \bibinfo {author} {\bibfnamefont {F.}~\bibnamefont {Reinhard}}, \ and\ \bibinfo {author} {\bibfnamefont {P.}~\bibnamefont {Cappellaro}},\ }\bibfield  {title} {\enquote {\bibinfo {title} {Quantum sensing},}\ }\href {\doibase 10.1103/RevModPhys.89.035002} {\bibfield  {journal} {\bibinfo  {journal} {Rev. Mod. Phys.}\ }\textbf {\bibinfo {volume} {89}},\ \bibinfo {pages} {035002} (\bibinfo {year} {2017})}\BibitemShut {NoStop}%
\bibitem [{\citenamefont {Fiderer}\ and\ \citenamefont {Braun}(2018)}]{Fiderer2018}%
  \BibitemOpen
  \bibfield  {author} {\bibinfo {author} {\bibfnamefont {Lukas~J.}\ \bibnamefont {Fiderer}}\ and\ \bibinfo {author} {\bibfnamefont {Daniel}\ \bibnamefont {Braun}},\ }\bibfield  {title} {\enquote {\bibinfo {title} {Quantum metrology with quantum-chaotic sensors},}\ }\href {\doibase 10.1038/s41467-018-03623-z} {\bibfield  {journal} {\bibinfo  {journal} {Nature Commun.}\ }\textbf {\bibinfo {volume} {9}},\ \bibinfo {pages} {1351} (\bibinfo {year} {2018})}\BibitemShut {NoStop}%
\bibitem [{\citenamefont {Crawford}\ \emph {et~al.}(2021)\citenamefont {Crawford}, \citenamefont {Shugayev}, \citenamefont {Paudel}, \citenamefont {Lu}, \citenamefont {Syamlal}, \citenamefont {Ohodnicki}, \citenamefont {Chorpening}, \citenamefont {Gentry},\ and\ \citenamefont {Duan}}]{Crawford2021}%
  \BibitemOpen
  \bibfield  {author} {\bibinfo {author} {\bibfnamefont {Scott~E.}\ \bibnamefont {Crawford}}, \bibinfo {author} {\bibfnamefont {Roman~A.}\ \bibnamefont {Shugayev}}, \bibinfo {author} {\bibfnamefont {Hari~P.}\ \bibnamefont {Paudel}}, \bibinfo {author} {\bibfnamefont {Ping}\ \bibnamefont {Lu}}, \bibinfo {author} {\bibfnamefont {Madhava}\ \bibnamefont {Syamlal}}, \bibinfo {author} {\bibfnamefont {Paul~R.}\ \bibnamefont {Ohodnicki}}, \bibinfo {author} {\bibfnamefont {Benjamin}\ \bibnamefont {Chorpening}}, \bibinfo {author} {\bibfnamefont {Randall}\ \bibnamefont {Gentry}}, \ and\ \bibinfo {author} {\bibfnamefont {Yuhua}\ \bibnamefont {Duan}},\ }\bibfield  {title} {\enquote {\bibinfo {title} {Quantum sensing for energy applications: Review and perspective},}\ }\href {\doibase https://doi.org/10.1002/qute.202100049} {\bibfield  {journal} {\bibinfo  {journal} {Adv. Quantum Technol.}\ }\textbf {\bibinfo {volume} {4}},\ \bibinfo {pages} {2100049} (\bibinfo {year} {2021})}\BibitemShut {NoStop}%
\bibitem [{\citenamefont {Montenegro}\ \emph {et~al.}(2025)\citenamefont {Montenegro}, \citenamefont {Mukhopadhyay}, \citenamefont {Yousefjani}, \citenamefont {Sarkar}, \citenamefont {Mishra}, \citenamefont {Paris},\ and\ \citenamefont {Bayat}}]{Montenegro2025}%
  \BibitemOpen
  \bibfield  {author} {\bibinfo {author} {\bibfnamefont {Victor}\ \bibnamefont {Montenegro}}, \bibinfo {author} {\bibfnamefont {Chiranjib}\ \bibnamefont {Mukhopadhyay}}, \bibinfo {author} {\bibfnamefont {Rozhin}\ \bibnamefont {Yousefjani}}, \bibinfo {author} {\bibfnamefont {Saubhik}\ \bibnamefont {Sarkar}}, \bibinfo {author} {\bibfnamefont {Utkarsh}\ \bibnamefont {Mishra}}, \bibinfo {author} {\bibfnamefont {Matteo~G.A.}\ \bibnamefont {Paris}}, \ and\ \bibinfo {author} {\bibfnamefont {Abolfazl}\ \bibnamefont {Bayat}},\ }\bibfield  {title} {\enquote {\bibinfo {title} {Review: Quantum metrology and sensing with many-body systems},}\ }\href {\doibase https://doi.org/10.1016/j.physrep.2025.05.005} {\bibfield  {journal} {\bibinfo  {journal} {Phys. Rep.}\ }\textbf {\bibinfo {volume} {1134}},\ \bibinfo {pages} {1--62} (\bibinfo {year} {2025})}\BibitemShut {NoStop}%
\bibitem [{\citenamefont {Lloyd}(1996)}]{Lloyd1996}%
  \BibitemOpen
  \bibfield  {author} {\bibinfo {author} {\bibfnamefont {Seth}\ \bibnamefont {Lloyd}},\ }\bibfield  {title} {\enquote {\bibinfo {title} {Universal quantum simulators},}\ }\href {\doibase 10.1126/science.273.5278.1073} {\bibfield  {journal} {\bibinfo  {journal} {Science}\ }\textbf {\bibinfo {volume} {273}},\ \bibinfo {pages} {1073--1078} (\bibinfo {year} {1996})}\BibitemShut {NoStop}%
\bibitem [{\citenamefont {Georgescu}\ \emph {et~al.}(2014)\citenamefont {Georgescu}, \citenamefont {Ashhab},\ and\ \citenamefont {Nori}}]{Georgescu2014}%
  \BibitemOpen
  \bibfield  {author} {\bibinfo {author} {\bibfnamefont {I.~M.}\ \bibnamefont {Georgescu}}, \bibinfo {author} {\bibfnamefont {S.}~\bibnamefont {Ashhab}}, \ and\ \bibinfo {author} {\bibfnamefont {Franco}\ \bibnamefont {Nori}},\ }\bibfield  {title} {\enquote {\bibinfo {title} {Quantum simulation},}\ }\href {\doibase 10.1103/RevModPhys.86.153} {\bibfield  {journal} {\bibinfo  {journal} {Rev. Mod. Phys.}\ }\textbf {\bibinfo {volume} {86}},\ \bibinfo {pages} {153--185} (\bibinfo {year} {2014})}\BibitemShut {NoStop}%
\bibitem [{\citenamefont {Zeilinger}(1999)}]{Zeilinger1999}%
  \BibitemOpen
  \bibfield  {author} {\bibinfo {author} {\bibfnamefont {Anton}\ \bibnamefont {Zeilinger}},\ }\bibfield  {title} {\enquote {\bibinfo {title} {Experiment and the foundations of quantum physics},}\ }\href {\doibase 10.1103/RevModPhys.71.S288} {\bibfield  {journal} {\bibinfo  {journal} {Rev. Mod. Phys.}\ }\textbf {\bibinfo {volume} {71}},\ \bibinfo {pages} {S288--S297} (\bibinfo {year} {1999})}\BibitemShut {NoStop}%
\bibitem [{\citenamefont {Gisin}\ and\ \citenamefont {Thew}(2007)}]{Gisin2007}%
  \BibitemOpen
  \bibfield  {author} {\bibinfo {author} {\bibfnamefont {Nicolas}\ \bibnamefont {Gisin}}\ and\ \bibinfo {author} {\bibfnamefont {Rob}\ \bibnamefont {Thew}},\ }\bibfield  {title} {\enquote {\bibinfo {title} {Quantum communication},}\ }\href {\doibase 10.1038/nphoton.2007.22} {\bibfield  {journal} {\bibinfo  {journal} {Nature Photonics}\ }\textbf {\bibinfo {volume} {1}},\ \bibinfo {pages} {165--171} (\bibinfo {year} {2007})}\BibitemShut {NoStop}%
\bibitem [{\citenamefont {Swingle}\ \emph {et~al.}(2016)\citenamefont {Swingle}, \citenamefont {Bentsen}, \citenamefont {Schleier-Smith},\ and\ \citenamefont {Hayden}}]{Swingle2016}%
  \BibitemOpen
  \bibfield  {author} {\bibinfo {author} {\bibfnamefont {Brian}\ \bibnamefont {Swingle}}, \bibinfo {author} {\bibfnamefont {Gregory}\ \bibnamefont {Bentsen}}, \bibinfo {author} {\bibfnamefont {Monika}\ \bibnamefont {Schleier-Smith}}, \ and\ \bibinfo {author} {\bibfnamefont {Patrick}\ \bibnamefont {Hayden}},\ }\bibfield  {title} {\enquote {\bibinfo {title} {Measuring the scrambling of quantum information},}\ }\href {\doibase 10.1103/PhysRevA.94.040302} {\bibfield  {journal} {\bibinfo  {journal} {Phys. Rev. A}\ }\textbf {\bibinfo {volume} {94}},\ \bibinfo {pages} {040302} (\bibinfo {year} {2016})}\BibitemShut {NoStop}%
\bibitem [{\citenamefont {Landsman}\ \emph {et~al.}(2019)\citenamefont {Landsman}, \citenamefont {Figgatt}, \citenamefont {Schuster}, \citenamefont {Linke}, \citenamefont {Yoshida}, \citenamefont {Yao},\ and\ \citenamefont {Monroe}}]{Landsman2019}%
  \BibitemOpen
  \bibfield  {author} {\bibinfo {author} {\bibfnamefont {K.~A.}\ \bibnamefont {Landsman}}, \bibinfo {author} {\bibfnamefont {C.}~\bibnamefont {Figgatt}}, \bibinfo {author} {\bibfnamefont {T.}~\bibnamefont {Schuster}}, \bibinfo {author} {\bibfnamefont {N.~M.}\ \bibnamefont {Linke}}, \bibinfo {author} {\bibfnamefont {B.}~\bibnamefont {Yoshida}}, \bibinfo {author} {\bibfnamefont {N.~Y.}\ \bibnamefont {Yao}}, \ and\ \bibinfo {author} {\bibfnamefont {C.}~\bibnamefont {Monroe}},\ }\bibfield  {title} {\enquote {\bibinfo {title} {Verified quantum information scrambling},}\ }\href {\doibase 10.1038/s41586-019-0952-6} {\bibfield  {journal} {\bibinfo  {journal} {Nature}\ }\textbf {\bibinfo {volume} {567}},\ \bibinfo {pages} {61--65} (\bibinfo {year} {2019})}\BibitemShut {NoStop}%
\bibitem [{\citenamefont {Dowling}\ and\ \citenamefont {Milburn}(2003)}]{Dowling2003}%
  \BibitemOpen
  \bibfield  {author} {\bibinfo {author} {\bibfnamefont {Jonathan~P.}\ \bibnamefont {Dowling}}\ and\ \bibinfo {author} {\bibfnamefont {Gerard~J.}\ \bibnamefont {Milburn}},\ }\bibfield  {title} {\enquote {\bibinfo {title} {Quantum technology: the second quantum revolution},}\ }\href {\doibase 10.1098/rsta.2003.1227} {\bibfield  {journal} {\bibinfo  {journal} {Phil. Trans. Roy. Soc. A}\ }\textbf {\bibinfo {volume} {361}},\ \bibinfo {pages} {1655--1674} (\bibinfo {year} {2003})}\BibitemShut {NoStop}%
\bibitem [{\citenamefont {Dicke}(1954)}]{Dicke1954}%
  \BibitemOpen
  \bibfield  {author} {\bibinfo {author} {\bibfnamefont {R.~H.}\ \bibnamefont {Dicke}},\ }\bibfield  {title} {\enquote {\bibinfo {title} {Coherence in spontaneous radiation processes},}\ }\href {\doibase 10.1103/PhysRev.93.99} {\bibfield  {journal} {\bibinfo  {journal} {Phys. Rev.}\ }\textbf {\bibinfo {volume} {93}},\ \bibinfo {pages} {99} (\bibinfo {year} {1954})}\BibitemShut {NoStop}%
\bibitem [{\citenamefont {Kirton}\ \emph {et~al.}(2019)\citenamefont {Kirton}, \citenamefont {Roses}, \citenamefont {Keeling},\ and\ \citenamefont {Dalla~Torre}}]{Kirton2019}%
  \BibitemOpen
  \bibfield  {author} {\bibinfo {author} {\bibfnamefont {Peter}\ \bibnamefont {Kirton}}, \bibinfo {author} {\bibfnamefont {Mor~M.}\ \bibnamefont {Roses}}, \bibinfo {author} {\bibfnamefont {Jonathan}\ \bibnamefont {Keeling}}, \ and\ \bibinfo {author} {\bibfnamefont {Emanuele~G.}\ \bibnamefont {Dalla~Torre}},\ }\bibfield  {title} {\enquote {\bibinfo {title} {Introduction to the {D}icke model: From equilibrium to nonequilibrium, and vice versa},}\ }\href {\doibase https://doi.org/10.1002/qute.201800043} {\bibfield  {journal} {\bibinfo  {journal} {Adv. Quantum Technol.}\ }\textbf {\bibinfo {volume} {2}},\ \bibinfo {pages} {1800043} (\bibinfo {year} {2019})}\BibitemShut {NoStop}%
\bibitem [{\citenamefont {Roses}\ and\ \citenamefont {Dalla~Torre}(2020)}]{Roses2020}%
  \BibitemOpen
  \bibfield  {author} {\bibinfo {author} {\bibfnamefont {Mor~M.}\ \bibnamefont {Roses}}\ and\ \bibinfo {author} {\bibfnamefont {Emanuele~G.}\ \bibnamefont {Dalla~Torre}},\ }\bibfield  {title} {\enquote {\bibinfo {title} {Dicke model},}\ }\href {\doibase 10.1371/journal.pone.0235197} {\bibfield  {journal} {\bibinfo  {journal} {PLOS ONE}\ }\textbf {\bibinfo {volume} {15}},\ \bibinfo {pages} {1--8} (\bibinfo {year} {2020})}\BibitemShut {NoStop}%
\bibitem [{\citenamefont {Larson}\ and\ \citenamefont {Mavrogordatos}(2021)}]{Larson2021Book}%
  \BibitemOpen
  \bibfield  {author} {\bibinfo {author} {\bibfnamefont {Jonas}\ \bibnamefont {Larson}}\ and\ \bibinfo {author} {\bibfnamefont {Themistoklis}\ \bibnamefont {Mavrogordatos}},\ }\href {\doibase 10.1088/978-0-7503-3447-1} {\emph {\bibinfo {title} {The {J}aynes–{C}ummings Model and Its Descendants}}},\ 2053-2563\ (\bibinfo  {publisher} {IOP Publishing},\ \bibinfo {year} {2021})\BibitemShut {NoStop}%
\bibitem [{\citenamefont {Villase{\~n}or}\ \emph {et~al.}(2024)\citenamefont {Villase{\~n}or}, \citenamefont {Pilatowsky-Cameo}, \citenamefont {Ch{\'a}vez-Carlos}, \citenamefont {Bastarrachea-Magnani}, \citenamefont {Lerma-Hern{\'a}ndez}, \citenamefont {Santos},\ and\ \citenamefont {Hirsch}}]{Villasenor2024ARXIV}%
  \BibitemOpen
  \bibfield  {author} {\bibinfo {author} {\bibfnamefont {David}\ \bibnamefont {Villase{\~n}or}}, \bibinfo {author} {\bibfnamefont {Sa{\'u}l}\ \bibnamefont {Pilatowsky-Cameo}}, \bibinfo {author} {\bibfnamefont {Jorge}\ \bibnamefont {Ch{\'a}vez-Carlos}}, \bibinfo {author} {\bibfnamefont {Miguel~A.}\ \bibnamefont {Bastarrachea-Magnani}}, \bibinfo {author} {\bibfnamefont {Sergio}\ \bibnamefont {Lerma-Hern{\'a}ndez}}, \bibinfo {author} {\bibfnamefont {Lea~F.}\ \bibnamefont {Santos}}, \ and\ \bibinfo {author} {\bibfnamefont {Jorge~G.}\ \bibnamefont {Hirsch}},\ }\href@noop {} {\enquote {\bibinfo {title} {Classical and quantum properties of the spin-boson {D}icke model: Chaos, localization, and scarring},}\ } (\bibinfo {year} {2024}),\ \Eprint {http://arxiv.org/abs/2405.20381} {arXiv:2405.20381 [quant-ph]} \BibitemShut {NoStop}%
\bibitem [{\citenamefont {Emary}\ and\ \citenamefont {Brandes}(2003{\natexlab{a}})}]{Emary2003}%
  \BibitemOpen
  \bibfield  {author} {\bibinfo {author} {\bibfnamefont {Clive}\ \bibnamefont {Emary}}\ and\ \bibinfo {author} {\bibfnamefont {Tobias}\ \bibnamefont {Brandes}},\ }\bibfield  {title} {\enquote {\bibinfo {title} {Chaos and the quantum phase transition in the {D}icke model},}\ }\href {\doibase 10.1103/PhysRevE.67.066203} {\bibfield  {journal} {\bibinfo  {journal} {Phys. Rev. E}\ }\textbf {\bibinfo {volume} {67}},\ \bibinfo {pages} {066203} (\bibinfo {year} {2003}{\natexlab{a}})}\BibitemShut {NoStop}%
\bibitem [{\citenamefont {Emary}\ and\ \citenamefont {Brandes}(2003{\natexlab{b}})}]{Emary2003PRL}%
  \BibitemOpen
  \bibfield  {author} {\bibinfo {author} {\bibfnamefont {Clive}\ \bibnamefont {Emary}}\ and\ \bibinfo {author} {\bibfnamefont {Tobias}\ \bibnamefont {Brandes}},\ }\bibfield  {title} {\enquote {\bibinfo {title} {Quantum chaos triggered by precursors of a quantum phase transition: The {D}icke model},}\ }\href {\doibase 10.1103/PhysRevLett.90.044101} {\bibfield  {journal} {\bibinfo  {journal} {Phys. Rev. Lett.}\ }\textbf {\bibinfo {volume} {90}},\ \bibinfo {pages} {044101} (\bibinfo {year} {2003}{\natexlab{b}})}\BibitemShut {NoStop}%
\bibitem [{\citenamefont {Lambert}\ \emph {et~al.}(2004)\citenamefont {Lambert}, \citenamefont {Emary},\ and\ \citenamefont {Brandes}}]{Lambert2004}%
  \BibitemOpen
  \bibfield  {author} {\bibinfo {author} {\bibfnamefont {Neill}\ \bibnamefont {Lambert}}, \bibinfo {author} {\bibfnamefont {Clive}\ \bibnamefont {Emary}}, \ and\ \bibinfo {author} {\bibfnamefont {Tobias}\ \bibnamefont {Brandes}},\ }\bibfield  {title} {\enquote {\bibinfo {title} {Entanglement and the phase transition in single-mode superradiance},}\ }\href {\doibase 10.1103/PhysRevLett.92.073602} {\bibfield  {journal} {\bibinfo  {journal} {Phys. Rev. Lett.}\ }\textbf {\bibinfo {volume} {92}},\ \bibinfo {pages} {073602} (\bibinfo {year} {2004})}\BibitemShut {NoStop}%
\bibitem [{\citenamefont {Brandes}(2005)}]{Brandes2005}%
  \BibitemOpen
  \bibfield  {author} {\bibinfo {author} {\bibfnamefont {Tobias}\ \bibnamefont {Brandes}},\ }\bibfield  {title} {\enquote {\bibinfo {title} {Coherent and collective quantum optical effects in mesoscopic systems},}\ }\href {\doibase https://doi.org/10.1016/j.physrep.2004.12.002} {\bibfield  {journal} {\bibinfo  {journal} {Phys. Rep.}\ }\textbf {\bibinfo {volume} {408}},\ \bibinfo {pages} {315--474} (\bibinfo {year} {2005})}\BibitemShut {NoStop}%
\bibitem [{\citenamefont {Brandes}(2013)}]{Brandes2013}%
  \BibitemOpen
  \bibfield  {author} {\bibinfo {author} {\bibfnamefont {Tobias}\ \bibnamefont {Brandes}},\ }\bibfield  {title} {\enquote {\bibinfo {title} {Excited-state quantum phase transitions in {D}icke superradiance models},}\ }\href {\doibase 10.1103/PhysRevE.88.032133} {\bibfield  {journal} {\bibinfo  {journal} {Phys. Rev. E}\ }\textbf {\bibinfo {volume} {88}},\ \bibinfo {pages} {032133} (\bibinfo {year} {2013})}\BibitemShut {NoStop}%
\bibitem [{\citenamefont {Rehler}\ and\ \citenamefont {Eberly}(1971)}]{Rehler1971}%
  \BibitemOpen
  \bibfield  {author} {\bibinfo {author} {\bibfnamefont {Nicholas~E.}\ \bibnamefont {Rehler}}\ and\ \bibinfo {author} {\bibfnamefont {Joseph~H.}\ \bibnamefont {Eberly}},\ }\bibfield  {title} {\enquote {\bibinfo {title} {Superradiance},}\ }\href {\doibase 10.1103/PhysRevA.3.1735} {\bibfield  {journal} {\bibinfo  {journal} {Phys. Rev. A}\ }\textbf {\bibinfo {volume} {3}},\ \bibinfo {pages} {1735--1751} (\bibinfo {year} {1971})}\BibitemShut {NoStop}%
\bibitem [{\citenamefont {Bonifacio}\ \emph {et~al.}(1971)\citenamefont {Bonifacio}, \citenamefont {Schwendimann},\ and\ \citenamefont {Haake}}]{Bonifacio1971}%
  \BibitemOpen
  \bibfield  {author} {\bibinfo {author} {\bibfnamefont {R.}~\bibnamefont {Bonifacio}}, \bibinfo {author} {\bibfnamefont {P.}~\bibnamefont {Schwendimann}}, \ and\ \bibinfo {author} {\bibfnamefont {Fritz}\ \bibnamefont {Haake}},\ }\bibfield  {title} {\enquote {\bibinfo {title} {Quantum statistical theory of superradiance. {I}},}\ }\href {\doibase 10.1103/PhysRevA.4.302} {\bibfield  {journal} {\bibinfo  {journal} {Phys. Rev. A}\ }\textbf {\bibinfo {volume} {4}},\ \bibinfo {pages} {302--313} (\bibinfo {year} {1971})}\BibitemShut {NoStop}%
\bibitem [{\citenamefont {MacGillivray}\ and\ \citenamefont {Feld}(1976)}]{MacGillivray1976}%
  \BibitemOpen
  \bibfield  {author} {\bibinfo {author} {\bibfnamefont {J.~C.}\ \bibnamefont {MacGillivray}}\ and\ \bibinfo {author} {\bibfnamefont {M.~S.}\ \bibnamefont {Feld}},\ }\bibfield  {title} {\enquote {\bibinfo {title} {Theory of superradiance in an extended, optically thick medium},}\ }\href {\doibase 10.1103/PhysRevA.14.1169} {\bibfield  {journal} {\bibinfo  {journal} {Phys. Rev. A}\ }\textbf {\bibinfo {volume} {14}},\ \bibinfo {pages} {1169--1189} (\bibinfo {year} {1976})}\BibitemShut {NoStop}%
\bibitem [{\citenamefont {Gross}\ and\ \citenamefont {Haroche}(1982)}]{Gross1982}%
  \BibitemOpen
  \bibfield  {author} {\bibinfo {author} {\bibfnamefont {Michel}\ \bibnamefont {Gross}}\ and\ \bibinfo {author} {\bibfnamefont {Serge}\ \bibnamefont {Haroche}},\ }\bibfield  {title} {\enquote {\bibinfo {title} {Superradiance: An essay on the theory of collective spontaneous emission},}\ }\href {\doibase 10.1016/0370-1573(82)90102-8} {\bibfield  {journal} {\bibinfo  {journal} {Phys. Rep.}\ }\textbf {\bibinfo {volume} {93}},\ \bibinfo {pages} {301--396} (\bibinfo {year} {1982})}\BibitemShut {NoStop}%
\bibitem [{\citenamefont {Sutherland}\ and\ \citenamefont {Robicheaux}(2017)}]{Sutherland2017}%
  \BibitemOpen
  \bibfield  {author} {\bibinfo {author} {\bibfnamefont {R.~T.}\ \bibnamefont {Sutherland}}\ and\ \bibinfo {author} {\bibfnamefont {F.}~\bibnamefont {Robicheaux}},\ }\bibfield  {title} {\enquote {\bibinfo {title} {Superradiance in inverted multilevel atomic clouds},}\ }\href {\doibase 10.1103/PhysRevA.95.033839} {\bibfield  {journal} {\bibinfo  {journal} {Phys. Rev. A}\ }\textbf {\bibinfo {volume} {95}},\ \bibinfo {pages} {033839} (\bibinfo {year} {2017})}\BibitemShut {NoStop}%
\bibitem [{\citenamefont {Masson}\ \emph {et~al.}(2020)\citenamefont {Masson}, \citenamefont {Ferrier-Barbut}, \citenamefont {Orozco}, \citenamefont {Browaeys},\ and\ \citenamefont {Asenjo-Garcia}}]{Masson2020}%
  \BibitemOpen
  \bibfield  {author} {\bibinfo {author} {\bibfnamefont {Stuart~J.}\ \bibnamefont {Masson}}, \bibinfo {author} {\bibfnamefont {Igor}\ \bibnamefont {Ferrier-Barbut}}, \bibinfo {author} {\bibfnamefont {Luis~A.}\ \bibnamefont {Orozco}}, \bibinfo {author} {\bibfnamefont {Antoine}\ \bibnamefont {Browaeys}}, \ and\ \bibinfo {author} {\bibfnamefont {Ana}\ \bibnamefont {Asenjo-Garcia}},\ }\bibfield  {title} {\enquote {\bibinfo {title} {Many-body signatures of collective decay in atomic chains},}\ }\href {\doibase 10.1103/PhysRevLett.125.263601} {\bibfield  {journal} {\bibinfo  {journal} {Phys. Rev. Lett.}\ }\textbf {\bibinfo {volume} {125}},\ \bibinfo {pages} {263601} (\bibinfo {year} {2020})}\BibitemShut {NoStop}%
\bibitem [{\citenamefont {Masson}\ and\ \citenamefont {Asenjo-Garcia}(2022)}]{Masson2022}%
  \BibitemOpen
  \bibfield  {author} {\bibinfo {author} {\bibfnamefont {Stuart~J.}\ \bibnamefont {Masson}}\ and\ \bibinfo {author} {\bibfnamefont {Ana}\ \bibnamefont {Asenjo-Garcia}},\ }\bibfield  {title} {\enquote {\bibinfo {title} {Universality of {D}icke superradiance in arrays of quantum emitters},}\ }\href {\doibase 10.1038/s41467-022-29805-4} {\bibfield  {journal} {\bibinfo  {journal} {Nature Communications}\ }\textbf {\bibinfo {volume} {13}},\ \bibinfo {pages} {2285} (\bibinfo {year} {2022})}\BibitemShut {NoStop}%
\bibitem [{\citenamefont {Sierra}\ \emph {et~al.}(2022)\citenamefont {Sierra}, \citenamefont {Masson},\ and\ \citenamefont {Asenjo-Garcia}}]{Sierra2022}%
  \BibitemOpen
  \bibfield  {author} {\bibinfo {author} {\bibfnamefont {Eric}\ \bibnamefont {Sierra}}, \bibinfo {author} {\bibfnamefont {Stuart~J.}\ \bibnamefont {Masson}}, \ and\ \bibinfo {author} {\bibfnamefont {Ana}\ \bibnamefont {Asenjo-Garcia}},\ }\bibfield  {title} {\enquote {\bibinfo {title} {Dicke superradiance in ordered lattices: Dimensionality matters},}\ }\href {\doibase 10.1103/PhysRevResearch.4.023207} {\bibfield  {journal} {\bibinfo  {journal} {Phys. Rev. Res.}\ }\textbf {\bibinfo {volume} {4}},\ \bibinfo {pages} {023207} (\bibinfo {year} {2022})}\BibitemShut {NoStop}%
\bibitem [{\citenamefont {Pi\~neiro Orioli}\ \emph {et~al.}(2022)\citenamefont {Pi\~neiro Orioli}, \citenamefont {Thompson},\ and\ \citenamefont {Rey}}]{Pineiro2022}%
  \BibitemOpen
  \bibfield  {author} {\bibinfo {author} {\bibfnamefont {A.}~\bibnamefont {Pi\~neiro Orioli}}, \bibinfo {author} {\bibfnamefont {J.~K.}\ \bibnamefont {Thompson}}, \ and\ \bibinfo {author} {\bibfnamefont {A.~M.}\ \bibnamefont {Rey}},\ }\bibfield  {title} {\enquote {\bibinfo {title} {Emergent dark states from superradiant dynamics in multilevel atoms in a cavity},}\ }\href {\doibase 10.1103/PhysRevX.12.011054} {\bibfield  {journal} {\bibinfo  {journal} {Phys. Rev. X}\ }\textbf {\bibinfo {volume} {12}},\ \bibinfo {pages} {011054} (\bibinfo {year} {2022})}\BibitemShut {NoStop}%
\bibitem [{\citenamefont {Cardenas-Lopez}\ \emph {et~al.}(2023)\citenamefont {Cardenas-Lopez}, \citenamefont {Masson}, \citenamefont {Zager},\ and\ \citenamefont {Asenjo-Garcia}}]{Cardenas2023}%
  \BibitemOpen
  \bibfield  {author} {\bibinfo {author} {\bibfnamefont {Silvia}\ \bibnamefont {Cardenas-Lopez}}, \bibinfo {author} {\bibfnamefont {Stuart~J.}\ \bibnamefont {Masson}}, \bibinfo {author} {\bibfnamefont {Zoe}\ \bibnamefont {Zager}}, \ and\ \bibinfo {author} {\bibfnamefont {Ana}\ \bibnamefont {Asenjo-Garcia}},\ }\bibfield  {title} {\enquote {\bibinfo {title} {Many-body superradiance and dynamical mirror symmetry breaking in waveguide {QED}},}\ }\href {\doibase 10.1103/PhysRevLett.131.033605} {\bibfield  {journal} {\bibinfo  {journal} {Phys. Rev. Lett.}\ }\textbf {\bibinfo {volume} {131}},\ \bibinfo {pages} {033605} (\bibinfo {year} {2023})}\BibitemShut {NoStop}%
\bibitem [{\citenamefont {Masson}\ \emph {et~al.}(2024)\citenamefont {Masson}, \citenamefont {Covey}, \citenamefont {Will},\ and\ \citenamefont {Asenjo-Garcia}}]{Masson2024}%
  \BibitemOpen
  \bibfield  {author} {\bibinfo {author} {\bibfnamefont {Stuart~J.}\ \bibnamefont {Masson}}, \bibinfo {author} {\bibfnamefont {Jacob~P.}\ \bibnamefont {Covey}}, \bibinfo {author} {\bibfnamefont {Sebastian}\ \bibnamefont {Will}}, \ and\ \bibinfo {author} {\bibfnamefont {Ana}\ \bibnamefont {Asenjo-Garcia}},\ }\bibfield  {title} {\enquote {\bibinfo {title} {Dicke superradiance in ordered arrays of multilevel atoms},}\ }\href {\doibase 10.1103/PRXQuantum.5.010344} {\bibfield  {journal} {\bibinfo  {journal} {PRX Quantum}\ }\textbf {\bibinfo {volume} {5}},\ \bibinfo {pages} {010344} (\bibinfo {year} {2024})}\BibitemShut {NoStop}%
\bibitem [{\citenamefont {Rosario}\ \emph {et~al.}(2025)\citenamefont {Rosario}, \citenamefont {Solak}, \citenamefont {Cidrim}, \citenamefont {Bachelard},\ and\ \citenamefont {Schachenmayer}}]{Rosario2025}%
  \BibitemOpen
  \bibfield  {author} {\bibinfo {author} {\bibfnamefont {P.}~\bibnamefont {Rosario}}, \bibinfo {author} {\bibfnamefont {L.~O.~R.}\ \bibnamefont {Solak}}, \bibinfo {author} {\bibfnamefont {A.}~\bibnamefont {Cidrim}}, \bibinfo {author} {\bibfnamefont {R.}~\bibnamefont {Bachelard}}, \ and\ \bibinfo {author} {\bibfnamefont {J.}~\bibnamefont {Schachenmayer}},\ }\bibfield  {title} {\enquote {\bibinfo {title} {Unraveling {D}icke superradiant decay with separable coherent spin states},}\ }\href {\doibase 10.1103/xcxr-sm9c} {\bibfield  {journal} {\bibinfo  {journal} {Phys. Rev. Lett.}\ }\textbf {\bibinfo {volume} {135}},\ \bibinfo {pages} {133602} (\bibinfo {year} {2025})}\BibitemShut {NoStop}%
\bibitem [{\citenamefont {Skribanowitz}\ \emph {et~al.}(1973)\citenamefont {Skribanowitz}, \citenamefont {Herman}, \citenamefont {MacGillivray},\ and\ \citenamefont {Feld}}]{Skribanowitz1973}%
  \BibitemOpen
  \bibfield  {author} {\bibinfo {author} {\bibfnamefont {N.}~\bibnamefont {Skribanowitz}}, \bibinfo {author} {\bibfnamefont {I.~P.}\ \bibnamefont {Herman}}, \bibinfo {author} {\bibfnamefont {J.~C.}\ \bibnamefont {MacGillivray}}, \ and\ \bibinfo {author} {\bibfnamefont {M.~S.}\ \bibnamefont {Feld}},\ }\bibfield  {title} {\enquote {\bibinfo {title} {Observation of {D}icke superradiance in optically pumped {HF} gas},}\ }\href {\doibase 10.1103/PhysRevLett.30.309} {\bibfield  {journal} {\bibinfo  {journal} {Phys. Rev. Lett.}\ }\textbf {\bibinfo {volume} {30}},\ \bibinfo {pages} {309--312} (\bibinfo {year} {1973})}\BibitemShut {NoStop}%
\bibitem [{\citenamefont {Raimond}\ \emph {et~al.}(1982)\citenamefont {Raimond}, \citenamefont {Goy}, \citenamefont {Gross}, \citenamefont {Fabre},\ and\ \citenamefont {Haroche}}]{Raimond1982}%
  \BibitemOpen
  \bibfield  {author} {\bibinfo {author} {\bibfnamefont {J.~M.}\ \bibnamefont {Raimond}}, \bibinfo {author} {\bibfnamefont {P.}~\bibnamefont {Goy}}, \bibinfo {author} {\bibfnamefont {M.}~\bibnamefont {Gross}}, \bibinfo {author} {\bibfnamefont {C.}~\bibnamefont {Fabre}}, \ and\ \bibinfo {author} {\bibfnamefont {S.}~\bibnamefont {Haroche}},\ }\bibfield  {title} {\enquote {\bibinfo {title} {Collective absorption of blackbody radiation by {R}ydberg atoms in a cavity: An experiment on {B}ose statistics and {B}rownian motion},}\ }\href {\doibase 10.1103/PhysRevLett.49.117} {\bibfield  {journal} {\bibinfo  {journal} {Phys. Rev. Lett.}\ }\textbf {\bibinfo {volume} {49}},\ \bibinfo {pages} {117--120} (\bibinfo {year} {1982})}\BibitemShut {NoStop}%
\bibitem [{\citenamefont {Inouye}\ \emph {et~al.}(1999)\citenamefont {Inouye}, \citenamefont {Chikkatur}, \citenamefont {Stamper-Kurn}, \citenamefont {Stenger}, \citenamefont {Pritchard},\ and\ \citenamefont {Ketterle}}]{Inouye1999}%
  \BibitemOpen
  \bibfield  {author} {\bibinfo {author} {\bibfnamefont {S.}~\bibnamefont {Inouye}}, \bibinfo {author} {\bibfnamefont {A.~P.}\ \bibnamefont {Chikkatur}}, \bibinfo {author} {\bibfnamefont {D.~M.}\ \bibnamefont {Stamper-Kurn}}, \bibinfo {author} {\bibfnamefont {J.}~\bibnamefont {Stenger}}, \bibinfo {author} {\bibfnamefont {D.~E.}\ \bibnamefont {Pritchard}}, \ and\ \bibinfo {author} {\bibfnamefont {W.}~\bibnamefont {Ketterle}},\ }\bibfield  {title} {\enquote {\bibinfo {title} {Superradiant {R}ayleigh scattering from a {B}ose-{E}instein condensate},}\ }\href {\doibase 10.1126/science.285.5427.571} {\bibfield  {journal} {\bibinfo  {journal} {Science}\ }\textbf {\bibinfo {volume} {285}},\ \bibinfo {pages} {571--574} (\bibinfo {year} {1999})}\BibitemShut {NoStop}%
\bibitem [{\citenamefont {Slama}\ \emph {et~al.}(2007)\citenamefont {Slama}, \citenamefont {Bux}, \citenamefont {Krenz}, \citenamefont {Zimmermann},\ and\ \citenamefont {Courteille}}]{Slama2007}%
  \BibitemOpen
  \bibfield  {author} {\bibinfo {author} {\bibfnamefont {S.}~\bibnamefont {Slama}}, \bibinfo {author} {\bibfnamefont {S.}~\bibnamefont {Bux}}, \bibinfo {author} {\bibfnamefont {G.}~\bibnamefont {Krenz}}, \bibinfo {author} {\bibfnamefont {C.}~\bibnamefont {Zimmermann}}, \ and\ \bibinfo {author} {\bibfnamefont {Ph.~W.}\ \bibnamefont {Courteille}},\ }\bibfield  {title} {\enquote {\bibinfo {title} {Superradiant {R}ayleigh scattering and collective atomic recoil lasing in a ring cavity},}\ }\href {\doibase 10.1103/PhysRevLett.98.053603} {\bibfield  {journal} {\bibinfo  {journal} {Phys. Rev. Lett.}\ }\textbf {\bibinfo {volume} {98}},\ \bibinfo {pages} {053603} (\bibinfo {year} {2007})}\BibitemShut {NoStop}%
\bibitem [{\citenamefont {Scheibner}\ \emph {et~al.}(2007)\citenamefont {Scheibner}, \citenamefont {Schmidt}, \citenamefont {Worschech}, \citenamefont {Forchel}, \citenamefont {Bacher}, \citenamefont {Passow},\ and\ \citenamefont {Hommel}}]{Scheibner2007}%
  \BibitemOpen
  \bibfield  {author} {\bibinfo {author} {\bibfnamefont {Michael}\ \bibnamefont {Scheibner}}, \bibinfo {author} {\bibfnamefont {Thomas}\ \bibnamefont {Schmidt}}, \bibinfo {author} {\bibfnamefont {Lukas}\ \bibnamefont {Worschech}}, \bibinfo {author} {\bibfnamefont {Alfred}\ \bibnamefont {Forchel}}, \bibinfo {author} {\bibfnamefont {Gerd}\ \bibnamefont {Bacher}}, \bibinfo {author} {\bibfnamefont {Thorsten}\ \bibnamefont {Passow}}, \ and\ \bibinfo {author} {\bibfnamefont {Detlef}\ \bibnamefont {Hommel}},\ }\bibfield  {title} {\enquote {\bibinfo {title} {Superradiance of quantum dots},}\ }\href {\doibase 10.1038/nphys494} {\bibfield  {journal} {\bibinfo  {journal} {Nat. Phys.}\ }\textbf {\bibinfo {volume} {3}},\ \bibinfo {pages} {106--110} (\bibinfo {year} {2007})}\BibitemShut {NoStop}%
\bibitem [{\citenamefont {Lo}\ \emph {et~al.}(1998)\citenamefont {Lo}, \citenamefont {Liu},\ and\ \citenamefont {Ng}}]{Lo1998}%
  \BibitemOpen
  \bibfield  {author} {\bibinfo {author} {\bibfnamefont {C.~F.}\ \bibnamefont {Lo}}, \bibinfo {author} {\bibfnamefont {K.~L.}\ \bibnamefont {Liu}}, \ and\ \bibinfo {author} {\bibfnamefont {K.~M.}\ \bibnamefont {Ng}},\ }\bibfield  {title} {\enquote {\bibinfo {title} {The multiquantum {J}aynes-{C}ummings model with the counter-rotating terms},}\ }\href {\doibase 10.1209/epl/i1998-00544-3} {\bibfield  {journal} {\bibinfo  {journal} {Europhys. Lett.}\ }\textbf {\bibinfo {volume} {42}},\ \bibinfo {pages} {1} (\bibinfo {year} {1998})}\BibitemShut {NoStop}%
\bibitem [{\citenamefont {Braak}(2026)}]{Braak2026}%
  \BibitemOpen
  \bibfield  {author} {\bibinfo {author} {\bibfnamefont {Daniel}\ \bibnamefont {Braak}},\ }\enquote {\bibinfo {title} {The $k$-photon quantum {R}abi model},}\ in\ \href {\doibase 10.1007/978-981-96-1218-5_5} {\emph {\bibinfo {booktitle} {Mathematical Foundations for Post-Quantum Cryptography: Crypto-Math CREST}}},\ \bibinfo {editor} {edited by\ \bibinfo {editor} {\bibfnamefont {Tsuyoshi}\ \bibnamefont {Takagi}}, \bibinfo {editor} {\bibfnamefont {Masato}\ \bibnamefont {Wakayama}}, \bibinfo {editor} {\bibfnamefont {Noboru}\ \bibnamefont {Kunihiro}}, \bibinfo {editor} {\bibfnamefont {Keisuke}\ \bibnamefont {Tanaka}}, \bibinfo {editor} {\bibfnamefont {Kazufumi}\ \bibnamefont {Kimoto}}, \ and\ \bibinfo {editor} {\bibfnamefont {Momonari}\ \bibnamefont {Kudo}}}\ (\bibinfo  {publisher} {Springer Nature Singapore},\ \bibinfo {address} {Singapore},\ \bibinfo {year} {2026})\ pp.\ \bibinfo {pages} {75--87}\BibitemShut {NoStop}%
\bibitem [{\citenamefont {Loy}(1978)}]{Loy1978}%
  \BibitemOpen
  \bibfield  {author} {\bibinfo {author} {\bibfnamefont {M.~M.~T.}\ \bibnamefont {Loy}},\ }\bibfield  {title} {\enquote {\bibinfo {title} {Two-photon adiabatic inversion},}\ }\href {\doibase 10.1103/PhysRevLett.41.473} {\bibfield  {journal} {\bibinfo  {journal} {Phys. Rev. Lett.}\ }\textbf {\bibinfo {volume} {41}},\ \bibinfo {pages} {473--476} (\bibinfo {year} {1978})}\BibitemShut {NoStop}%
\bibitem [{\citenamefont {Schlemmer}\ \emph {et~al.}(1980)\citenamefont {Schlemmer}, \citenamefont {Frölich},\ and\ \citenamefont {Welling}}]{Schlemmer1980}%
  \BibitemOpen
  \bibfield  {author} {\bibinfo {author} {\bibfnamefont {H.}~\bibnamefont {Schlemmer}}, \bibinfo {author} {\bibfnamefont {D.}~\bibnamefont {Frölich}}, \ and\ \bibinfo {author} {\bibfnamefont {H.}~\bibnamefont {Welling}},\ }\bibfield  {title} {\enquote {\bibinfo {title} {Two-photon amplification on cascade-transitions},}\ }\href {\doibase https://doi.org/10.1016/0030-4018(80)90333-8} {\bibfield  {journal} {\bibinfo  {journal} {Opt. Commun.}\ }\textbf {\bibinfo {volume} {32}},\ \bibinfo {pages} {141--144} (\bibinfo {year} {1980})}\BibitemShut {NoStop}%
\bibitem [{\citenamefont {Nikolaus}\ \emph {et~al.}(1981)\citenamefont {Nikolaus}, \citenamefont {Zhang},\ and\ \citenamefont {Toschek}}]{Nikolaus1981}%
  \BibitemOpen
  \bibfield  {author} {\bibinfo {author} {\bibfnamefont {B.}~\bibnamefont {Nikolaus}}, \bibinfo {author} {\bibfnamefont {D.~Z.}\ \bibnamefont {Zhang}}, \ and\ \bibinfo {author} {\bibfnamefont {P.~E.}\ \bibnamefont {Toschek}},\ }\bibfield  {title} {\enquote {\bibinfo {title} {Two-photon laser},}\ }\href {\doibase 10.1103/PhysRevLett.47.171} {\bibfield  {journal} {\bibinfo  {journal} {Phys. Rev. Lett.}\ }\textbf {\bibinfo {volume} {47}},\ \bibinfo {pages} {171--173} (\bibinfo {year} {1981})}\BibitemShut {NoStop}%
\bibitem [{\citenamefont {Brune}\ \emph {et~al.}(1987{\natexlab{a}})\citenamefont {Brune}, \citenamefont {Raimond},\ and\ \citenamefont {Haroche}}]{Brune1987}%
  \BibitemOpen
  \bibfield  {author} {\bibinfo {author} {\bibfnamefont {M.}~\bibnamefont {Brune}}, \bibinfo {author} {\bibfnamefont {J.~M.}\ \bibnamefont {Raimond}}, \ and\ \bibinfo {author} {\bibfnamefont {S.}~\bibnamefont {Haroche}},\ }\bibfield  {title} {\enquote {\bibinfo {title} {Theory of the {R}ydberg-atom two-photon micromaser},}\ }\href {\doibase 10.1103/PhysRevA.35.154} {\bibfield  {journal} {\bibinfo  {journal} {Phys. Rev. A}\ }\textbf {\bibinfo {volume} {35}},\ \bibinfo {pages} {154--163} (\bibinfo {year} {1987}{\natexlab{a}})}\BibitemShut {NoStop}%
\bibitem [{\citenamefont {Brune}\ \emph {et~al.}(1987{\natexlab{b}})\citenamefont {Brune}, \citenamefont {Raimond}, \citenamefont {Goy}, \citenamefont {Davidovich},\ and\ \citenamefont {Haroche}}]{Brune1987PRL}%
  \BibitemOpen
  \bibfield  {author} {\bibinfo {author} {\bibfnamefont {M.}~\bibnamefont {Brune}}, \bibinfo {author} {\bibfnamefont {J.~M.}\ \bibnamefont {Raimond}}, \bibinfo {author} {\bibfnamefont {P.}~\bibnamefont {Goy}}, \bibinfo {author} {\bibfnamefont {L.}~\bibnamefont {Davidovich}}, \ and\ \bibinfo {author} {\bibfnamefont {S.}~\bibnamefont {Haroche}},\ }\bibfield  {title} {\enquote {\bibinfo {title} {Realization of a two-photon maser oscillator},}\ }\href {\doibase 10.1103/PhysRevLett.59.1899} {\bibfield  {journal} {\bibinfo  {journal} {Phys. Rev. Lett.}\ }\textbf {\bibinfo {volume} {59}},\ \bibinfo {pages} {1899--1902} (\bibinfo {year} {1987}{\natexlab{b}})}\BibitemShut {NoStop}%
\bibitem [{\citenamefont {Brune}\ \emph {et~al.}(1988)\citenamefont {Brune}, \citenamefont {Raimond}, \citenamefont {Goy}, \citenamefont {Davidovich},\ and\ \citenamefont {Haroche}}]{Brune1988}%
  \BibitemOpen
  \bibfield  {author} {\bibinfo {author} {\bibfnamefont {M.}~\bibnamefont {Brune}}, \bibinfo {author} {\bibfnamefont {J.M.}\ \bibnamefont {Raimond}}, \bibinfo {author} {\bibfnamefont {P.}~\bibnamefont {Goy}}, \bibinfo {author} {\bibfnamefont {L.}~\bibnamefont {Davidovich}}, \ and\ \bibinfo {author} {\bibfnamefont {S.}~\bibnamefont {Haroche}},\ }\bibfield  {title} {\enquote {\bibinfo {title} {The two-photon {R}ydberg atom micromaser},}\ }\href {\doibase 10.1109/3.970} {\bibfield  {journal} {\bibinfo  {journal} {IEEE J. Quantum Electronics}\ }\textbf {\bibinfo {volume} {24}},\ \bibinfo {pages} {1323--1330} (\bibinfo {year} {1988})}\BibitemShut {NoStop}%
\bibitem [{\citenamefont {Gauthier}\ \emph {et~al.}(1992)\citenamefont {Gauthier}, \citenamefont {Wu}, \citenamefont {Morin},\ and\ \citenamefont {Mossberg}}]{Gauthier1992}%
  \BibitemOpen
  \bibfield  {author} {\bibinfo {author} {\bibfnamefont {Daniel~J.}\ \bibnamefont {Gauthier}}, \bibinfo {author} {\bibfnamefont {Qilin}\ \bibnamefont {Wu}}, \bibinfo {author} {\bibfnamefont {S.~E.}\ \bibnamefont {Morin}}, \ and\ \bibinfo {author} {\bibfnamefont {T.~W.}\ \bibnamefont {Mossberg}},\ }\bibfield  {title} {\enquote {\bibinfo {title} {Realization of a continuous-wave, two-photon optical laser},}\ }\href {\doibase 10.1103/PhysRevLett.68.464} {\bibfield  {journal} {\bibinfo  {journal} {Phys. Rev. Lett.}\ }\textbf {\bibinfo {volume} {68}},\ \bibinfo {pages} {464--467} (\bibinfo {year} {1992})}\BibitemShut {NoStop}%
\bibitem [{\citenamefont {Stufler}\ \emph {et~al.}(2006)\citenamefont {Stufler}, \citenamefont {Machnikowski}, \citenamefont {Ester}, \citenamefont {Bichler}, \citenamefont {Axt}, \citenamefont {Kuhn},\ and\ \citenamefont {Zrenner}}]{Stufler2006}%
  \BibitemOpen
  \bibfield  {author} {\bibinfo {author} {\bibfnamefont {S.}~\bibnamefont {Stufler}}, \bibinfo {author} {\bibfnamefont {P.}~\bibnamefont {Machnikowski}}, \bibinfo {author} {\bibfnamefont {P.}~\bibnamefont {Ester}}, \bibinfo {author} {\bibfnamefont {M.}~\bibnamefont {Bichler}}, \bibinfo {author} {\bibfnamefont {V.~M.}\ \bibnamefont {Axt}}, \bibinfo {author} {\bibfnamefont {T.}~\bibnamefont {Kuhn}}, \ and\ \bibinfo {author} {\bibfnamefont {A.}~\bibnamefont {Zrenner}},\ }\bibfield  {title} {\enquote {\bibinfo {title} {Two-photon {R}abi oscillations in a single $\text{In}_{x}\text{Ga}_{1-x}${As/GaAs} quantum dot},}\ }\href {\doibase 10.1103/PhysRevB.73.125304} {\bibfield  {journal} {\bibinfo  {journal} {Phys. Rev. B}\ }\textbf {\bibinfo {volume} {73}},\ \bibinfo {pages} {125304} (\bibinfo {year} {2006})}\BibitemShut {NoStop}%
\bibitem [{\citenamefont {del Valle}\ \emph {et~al.}(2010)\citenamefont {del Valle}, \citenamefont {Zippilli}, \citenamefont {Laussy}, \citenamefont {Gonzalez-Tudela}, \citenamefont {Morigi},\ and\ \citenamefont {Tejedor}}]{DelValle2010}%
  \BibitemOpen
  \bibfield  {author} {\bibinfo {author} {\bibfnamefont {Elena}\ \bibnamefont {del Valle}}, \bibinfo {author} {\bibfnamefont {Stefano}\ \bibnamefont {Zippilli}}, \bibinfo {author} {\bibfnamefont {Fabrice~P.}\ \bibnamefont {Laussy}}, \bibinfo {author} {\bibfnamefont {Alejandro}\ \bibnamefont {Gonzalez-Tudela}}, \bibinfo {author} {\bibfnamefont {Giovanna}\ \bibnamefont {Morigi}}, \ and\ \bibinfo {author} {\bibfnamefont {Carlos}\ \bibnamefont {Tejedor}},\ }\bibfield  {title} {\enquote {\bibinfo {title} {Two-photon lasing by a single quantum dot in a high-$q$ microcavity},}\ }\href {\doibase 10.1103/PhysRevB.81.035302} {\bibfield  {journal} {\bibinfo  {journal} {Phys. Rev. B}\ }\textbf {\bibinfo {volume} {81}},\ \bibinfo {pages} {035302} (\bibinfo {year} {2010})}\BibitemShut {NoStop}%
\bibitem [{\citenamefont {Felicetti}\ \emph {et~al.}(2018)\citenamefont {Felicetti}, \citenamefont {Rossatto}, \citenamefont {Rico}, \citenamefont {Solano},\ and\ \citenamefont {Forn-D\'{\i}az}}]{Felicetti2018}%
  \BibitemOpen
  \bibfield  {author} {\bibinfo {author} {\bibfnamefont {S.}~\bibnamefont {Felicetti}}, \bibinfo {author} {\bibfnamefont {D.~Z.}\ \bibnamefont {Rossatto}}, \bibinfo {author} {\bibfnamefont {E.}~\bibnamefont {Rico}}, \bibinfo {author} {\bibfnamefont {E.}~\bibnamefont {Solano}}, \ and\ \bibinfo {author} {\bibfnamefont {P.}~\bibnamefont {Forn-D\'{\i}az}},\ }\bibfield  {title} {\enquote {\bibinfo {title} {Two-photon quantum {R}abi model with superconducting circuits},}\ }\href {\doibase 10.1103/PhysRevA.97.013851} {\bibfield  {journal} {\bibinfo  {journal} {Phys. Rev. A}\ }\textbf {\bibinfo {volume} {97}},\ \bibinfo {pages} {013851} (\bibinfo {year} {2018})}\BibitemShut {NoStop}%
\bibitem [{\citenamefont {Crescente}\ \emph {et~al.}(2020)\citenamefont {Crescente}, \citenamefont {Carrega}, \citenamefont {Sassetti},\ and\ \citenamefont {Ferraro}}]{Crescente2020}%
  \BibitemOpen
  \bibfield  {author} {\bibinfo {author} {\bibfnamefont {Alba}\ \bibnamefont {Crescente}}, \bibinfo {author} {\bibfnamefont {Matteo}\ \bibnamefont {Carrega}}, \bibinfo {author} {\bibfnamefont {Maura}\ \bibnamefont {Sassetti}}, \ and\ \bibinfo {author} {\bibfnamefont {Dario}\ \bibnamefont {Ferraro}},\ }\bibfield  {title} {\enquote {\bibinfo {title} {Ultrafast charging in a two-photon {D}icke quantum battery},}\ }\href {\doibase 10.1103/PhysRevB.102.245407} {\bibfield  {journal} {\bibinfo  {journal} {Phys. Rev. B}\ }\textbf {\bibinfo {volume} {102}},\ \bibinfo {pages} {245407} (\bibinfo {year} {2020})}\BibitemShut {NoStop}%
\bibitem [{\citenamefont {Wang}\ \emph {et~al.}(2024)\citenamefont {Wang}, \citenamefont {Liu}, \citenamefont {Wu}, \citenamefont {Fan},\ and\ \citenamefont {Liu}}]{Wang2024}%
  \BibitemOpen
  \bibfield  {author} {\bibinfo {author} {\bibfnamefont {Lu}~\bibnamefont {Wang}}, \bibinfo {author} {\bibfnamefont {Shu-Qian}\ \bibnamefont {Liu}}, \bibinfo {author} {\bibfnamefont {Feng-lin}\ \bibnamefont {Wu}}, \bibinfo {author} {\bibfnamefont {Hao}\ \bibnamefont {Fan}}, \ and\ \bibinfo {author} {\bibfnamefont {Si-Yuan}\ \bibnamefont {Liu}},\ }\bibfield  {title} {\enquote {\bibinfo {title} {Deep strong charging in a multiphoton anisotropic {D}icke quantum battery},}\ }\href {\doibase 10.1103/PhysRevA.110.042419} {\bibfield  {journal} {\bibinfo  {journal} {Phys. Rev. A}\ }\textbf {\bibinfo {volume} {110}},\ \bibinfo {pages} {042419} (\bibinfo {year} {2024})}\BibitemShut {NoStop}%
\bibitem [{\citenamefont {Agarwal}(1985)}]{Agarwal1985}%
  \BibitemOpen
  \bibfield  {author} {\bibinfo {author} {\bibfnamefont {G.~S.}\ \bibnamefont {Agarwal}},\ }\bibfield  {title} {\enquote {\bibinfo {title} {Vacuum-field {R}abi oscillations of atoms in a cavity},}\ }\href {\doibase 10.1364/JOSAB.2.000480} {\bibfield  {journal} {\bibinfo  {journal} {J. Opt. Soc. Am. B}\ }\textbf {\bibinfo {volume} {2}},\ \bibinfo {pages} {480--485} (\bibinfo {year} {1985})}\BibitemShut {NoStop}%
\bibitem [{\citenamefont {Alsing}\ and\ \citenamefont {Zubairy}(1987)}]{Alsing1987}%
  \BibitemOpen
  \bibfield  {author} {\bibinfo {author} {\bibfnamefont {P.}~\bibnamefont {Alsing}}\ and\ \bibinfo {author} {\bibfnamefont {M.~S.}\ \bibnamefont {Zubairy}},\ }\bibfield  {title} {\enquote {\bibinfo {title} {Collapse and revivals in a two-photon absorption process},}\ }\href {\doibase 10.1364/JOSAB.4.000177} {\bibfield  {journal} {\bibinfo  {journal} {J. Opt. Soc. Am. B}\ }\textbf {\bibinfo {volume} {4}},\ \bibinfo {pages} {177--184} (\bibinfo {year} {1987})}\BibitemShut {NoStop}%
\bibitem [{\citenamefont {Puri}\ and\ \citenamefont {Bullough}(1988)}]{Puri1988}%
  \BibitemOpen
  \bibfield  {author} {\bibinfo {author} {\bibfnamefont {R.~R.}\ \bibnamefont {Puri}}\ and\ \bibinfo {author} {\bibfnamefont {R.~K.}\ \bibnamefont {Bullough}},\ }\bibfield  {title} {\enquote {\bibinfo {title} {Quantum electrodynamics of an atom making two-photon transitions in an ideal cavity},}\ }\href {\doibase 10.1364/JOSAB.5.002021} {\bibfield  {journal} {\bibinfo  {journal} {J. Opt. Soc. Am. B}\ }\textbf {\bibinfo {volume} {5}},\ \bibinfo {pages} {2021--2028} (\bibinfo {year} {1988})}\BibitemShut {NoStop}%
\bibitem [{\citenamefont {Toor}\ and\ \citenamefont {Zubairy}(1992)}]{Toor1992}%
  \BibitemOpen
  \bibfield  {author} {\bibinfo {author} {\bibfnamefont {A.~H.}\ \bibnamefont {Toor}}\ and\ \bibinfo {author} {\bibfnamefont {M.~S.}\ \bibnamefont {Zubairy}},\ }\bibfield  {title} {\enquote {\bibinfo {title} {Validity of the effective {H}amiltonian in the two-photon atom-field interaction},}\ }\href {\doibase 10.1103/PhysRevA.45.4951} {\bibfield  {journal} {\bibinfo  {journal} {Phys. Rev. A}\ }\textbf {\bibinfo {volume} {45}},\ \bibinfo {pages} {4951--4959} (\bibinfo {year} {1992})}\BibitemShut {NoStop}%
\bibitem [{\citenamefont {Puri}\ and\ \citenamefont {Agarwal}(1988)}]{Puri1988PRA}%
  \BibitemOpen
  \bibfield  {author} {\bibinfo {author} {\bibfnamefont {R.~R.}\ \bibnamefont {Puri}}\ and\ \bibinfo {author} {\bibfnamefont {G.~S.}\ \bibnamefont {Agarwal}},\ }\bibfield  {title} {\enquote {\bibinfo {title} {Coherent two-photon transitions in {R}ydberg atoms in a cavity with finite {Q}},}\ }\href {\doibase 10.1103/PhysRevA.37.3879} {\bibfield  {journal} {\bibinfo  {journal} {Phys. Rev. A}\ }\textbf {\bibinfo {volume} {37}},\ \bibinfo {pages} {3879--3883} (\bibinfo {year} {1988})}\BibitemShut {NoStop}%
\bibitem [{\citenamefont {Felicetti}\ \emph {et~al.}(2015)\citenamefont {Felicetti}, \citenamefont {Pedernales}, \citenamefont {Egusquiza}, \citenamefont {Romero}, \citenamefont {Lamata}, \citenamefont {Braak},\ and\ \citenamefont {Solano}}]{Felicetti2015}%
  \BibitemOpen
  \bibfield  {author} {\bibinfo {author} {\bibfnamefont {S.}~\bibnamefont {Felicetti}}, \bibinfo {author} {\bibfnamefont {J.~S.}\ \bibnamefont {Pedernales}}, \bibinfo {author} {\bibfnamefont {I.~L.}\ \bibnamefont {Egusquiza}}, \bibinfo {author} {\bibfnamefont {G.}~\bibnamefont {Romero}}, \bibinfo {author} {\bibfnamefont {L.}~\bibnamefont {Lamata}}, \bibinfo {author} {\bibfnamefont {D.}~\bibnamefont {Braak}}, \ and\ \bibinfo {author} {\bibfnamefont {E.}~\bibnamefont {Solano}},\ }\bibfield  {title} {\enquote {\bibinfo {title} {Spectral collapse via two-phonon interactions in trapped ions},}\ }\href {\doibase 10.1103/PhysRevA.92.033817} {\bibfield  {journal} {\bibinfo  {journal} {Phys. Rev. A}\ }\textbf {\bibinfo {volume} {92}},\ \bibinfo {pages} {033817} (\bibinfo {year} {2015})}\BibitemShut {NoStop}%
\bibitem [{\citenamefont {Duan}\ \emph {et~al.}(2016)\citenamefont {Duan}, \citenamefont {Xie}, \citenamefont {Braak},\ and\ \citenamefont {Chen}}]{Duan2016}%
  \BibitemOpen
  \bibfield  {author} {\bibinfo {author} {\bibfnamefont {Liwei}\ \bibnamefont {Duan}}, \bibinfo {author} {\bibfnamefont {You-Fei}\ \bibnamefont {Xie}}, \bibinfo {author} {\bibfnamefont {Daniel}\ \bibnamefont {Braak}}, \ and\ \bibinfo {author} {\bibfnamefont {Qing-Hu}\ \bibnamefont {Chen}},\ }\bibfield  {title} {\enquote {\bibinfo {title} {Two-photon {R}abi model: analytic solutions and spectral collapse},}\ }\href {\doibase 10.1088/1751-8113/49/46/464002} {\bibfield  {journal} {\bibinfo  {journal} {J. Phys. A: Math. Theor.}\ }\textbf {\bibinfo {volume} {49}},\ \bibinfo {pages} {464002} (\bibinfo {year} {2016})}\BibitemShut {NoStop}%
\bibitem [{\citenamefont {Rico}\ \emph {et~al.}(2020)\citenamefont {Rico}, \citenamefont {Maldonado-Villamizar},\ and\ \citenamefont {Rodriguez-Lara}}]{Rico2020}%
  \BibitemOpen
  \bibfield  {author} {\bibinfo {author} {\bibfnamefont {R.~J.~Armenta}\ \bibnamefont {Rico}}, \bibinfo {author} {\bibfnamefont {F.~H.}\ \bibnamefont {Maldonado-Villamizar}}, \ and\ \bibinfo {author} {\bibfnamefont {B.~M.}\ \bibnamefont {Rodriguez-Lara}},\ }\bibfield  {title} {\enquote {\bibinfo {title} {Spectral collapse in the two-photon quantum {R}abi model},}\ }\href {\doibase 10.1103/PhysRevA.101.063825} {\bibfield  {journal} {\bibinfo  {journal} {Phys. Rev. A}\ }\textbf {\bibinfo {volume} {101}},\ \bibinfo {pages} {063825} (\bibinfo {year} {2020})}\BibitemShut {NoStop}%
\bibitem [{\citenamefont {Lo}(2021)}]{Lo2021}%
  \BibitemOpen
  \bibfield  {author} {\bibinfo {author} {\bibfnamefont {C.~F.}\ \bibnamefont {Lo}},\ }\bibfield  {title} {\enquote {\bibinfo {title} {Spectral collapse in multiqubit two-photon {R}abi model},}\ }\href {\doibase 10.1038/s41598-021-84934-y} {\bibfield  {journal} {\bibinfo  {journal} {Sci. Rep.}\ }\textbf {\bibinfo {volume} {11}},\ \bibinfo {pages} {5409} (\bibinfo {year} {2021})}\BibitemShut {NoStop}%
\bibitem [{\citenamefont {Gerry}\ and\ \citenamefont {Togeas}(1989)}]{Gerry1989}%
  \BibitemOpen
  \bibfield  {author} {\bibinfo {author} {\bibfnamefont {Christopher~C.}\ \bibnamefont {Gerry}}\ and\ \bibinfo {author} {\bibfnamefont {James~B.}\ \bibnamefont {Togeas}},\ }\bibfield  {title} {\enquote {\bibinfo {title} {Squeezing and photon antibunching from a two-photon {D}icke model},}\ }\href {\doibase https://doi.org/10.1016/0030-4018(89)90112-0} {\bibfield  {journal} {\bibinfo  {journal} {Optics Communications}\ }\textbf {\bibinfo {volume} {69}},\ \bibinfo {pages} {263--266} (\bibinfo {year} {1989})}\BibitemShut {NoStop}%
\bibitem [{\citenamefont {Peng}\ \emph {et~al.}(2017)\citenamefont {Peng}, \citenamefont {Zheng}, \citenamefont {Guo}, \citenamefont {Guo}, \citenamefont {Zhang}, \citenamefont {Deng}, \citenamefont {Ju}, \citenamefont {Ren}, \citenamefont {Lamata},\ and\ \citenamefont {Solano}}]{Peng2017}%
  \BibitemOpen
  \bibfield  {author} {\bibinfo {author} {\bibfnamefont {Jie}\ \bibnamefont {Peng}}, \bibinfo {author} {\bibfnamefont {Chenxiong}\ \bibnamefont {Zheng}}, \bibinfo {author} {\bibfnamefont {Guangjie}\ \bibnamefont {Guo}}, \bibinfo {author} {\bibfnamefont {Xiaoyong}\ \bibnamefont {Guo}}, \bibinfo {author} {\bibfnamefont {Xin}\ \bibnamefont {Zhang}}, \bibinfo {author} {\bibfnamefont {Chaosheng}\ \bibnamefont {Deng}}, \bibinfo {author} {\bibfnamefont {Guoxing}\ \bibnamefont {Ju}}, \bibinfo {author} {\bibfnamefont {Zhongzhou}\ \bibnamefont {Ren}}, \bibinfo {author} {\bibfnamefont {Lucas}\ \bibnamefont {Lamata}}, \ and\ \bibinfo {author} {\bibfnamefont {Enrique}\ \bibnamefont {Solano}},\ }\bibfield  {title} {\enquote {\bibinfo {title} {Dark-like states for the multi-qubit and multi-photon {R}abi models},}\ }\href {\doibase 10.1088/1751-8121/aa651d} {\bibfield  {journal} {\bibinfo  {journal} {J. Phys. A: Math. Theor.}\ }\textbf {\bibinfo {volume} {50}},\ \bibinfo {pages} {174003} (\bibinfo {year} {2017})}\BibitemShut
  {NoStop}%
\bibitem [{\citenamefont {Wang}\ \emph {et~al.}(2019)\citenamefont {Wang}, \citenamefont {Chen},\ and\ \citenamefont {Jing}}]{Wang2019TPD}%
  \BibitemOpen
  \bibfield  {author} {\bibinfo {author} {\bibfnamefont {Shangyun}\ \bibnamefont {Wang}}, \bibinfo {author} {\bibfnamefont {Songbai}\ \bibnamefont {Chen}}, \ and\ \bibinfo {author} {\bibfnamefont {Jiliang}\ \bibnamefont {Jing}},\ }\bibfield  {title} {\enquote {\bibinfo {title} {Effect of system energy on quantum signatures of chaos in the two-photon {D}icke model},}\ }\href {\doibase 10.1103/PhysRevE.100.022207} {\bibfield  {journal} {\bibinfo  {journal} {Phys. Rev. E}\ }\textbf {\bibinfo {volume} {100}},\ \bibinfo {pages} {022207} (\bibinfo {year} {2019})}\BibitemShut {NoStop}%
\bibitem [{\citenamefont {Banerjee}\ \emph {et~al.}(2022)\citenamefont {Banerjee}, \citenamefont {Sharma},\ and\ \citenamefont {Bhattacherjee}}]{Banerjee2022}%
  \BibitemOpen
  \bibfield  {author} {\bibinfo {author} {\bibfnamefont {Priyankar}\ \bibnamefont {Banerjee}}, \bibinfo {author} {\bibfnamefont {Deepti}\ \bibnamefont {Sharma}}, \ and\ \bibinfo {author} {\bibfnamefont {Aranya~B.}\ \bibnamefont {Bhattacherjee}},\ }\bibfield  {title} {\enquote {\bibinfo {title} {Enhanced photon squeezing in two-photon {D}icke model},}\ }\href {\doibase https://doi.org/10.1016/j.physleta.2022.128287} {\bibfield  {journal} {\bibinfo  {journal} {Phys. Lett. A}\ }\textbf {\bibinfo {volume} {446}},\ \bibinfo {pages} {128287} (\bibinfo {year} {2022})}\BibitemShut {NoStop}%
\bibitem [{\citenamefont {Ram\'{\i}rez}\ \emph {et~al.}(2025)\citenamefont {Ram\'{\i}rez}, \citenamefont {Villase\~nor}, \citenamefont {Morales-Guzm\'an}, \citenamefont {Castro},\ and\ \citenamefont {Hirsch}}]{Ramirez2025}%
  \BibitemOpen
  \bibfield  {author} {\bibinfo {author} {\bibfnamefont {Fabrizio}\ \bibnamefont {Ram\'{\i}rez}}, \bibinfo {author} {\bibfnamefont {David}\ \bibnamefont {Villase\~nor}}, \bibinfo {author} {\bibfnamefont {Viani~S.}\ \bibnamefont {Morales-Guzm\'an}}, \bibinfo {author} {\bibfnamefont {Darly~Y.}\ \bibnamefont {Castro}}, \ and\ \bibinfo {author} {\bibfnamefont {Jorge~G.}\ \bibnamefont {Hirsch}},\ }\bibfield  {title} {\enquote {\bibinfo {title} {Loss of integrability in a system with two-photon interactions},}\ }\href {\doibase 10.1103/c618-hql5} {\bibfield  {journal} {\bibinfo  {journal} {Phys. Rev. A}\ }\textbf {\bibinfo {volume} {112}},\ \bibinfo {pages} {063707} (\bibinfo {year} {2025})}\BibitemShut {NoStop}%
\bibitem [{\citenamefont {Garbe}\ \emph {et~al.}(2017)\citenamefont {Garbe}, \citenamefont {Egusquiza}, \citenamefont {Solano}, \citenamefont {Ciuti}, \citenamefont {Coudreau}, \citenamefont {Milman},\ and\ \citenamefont {Felicetti}}]{Garbe2017}%
  \BibitemOpen
  \bibfield  {author} {\bibinfo {author} {\bibfnamefont {L.}~\bibnamefont {Garbe}}, \bibinfo {author} {\bibfnamefont {I.~L.}\ \bibnamefont {Egusquiza}}, \bibinfo {author} {\bibfnamefont {E.}~\bibnamefont {Solano}}, \bibinfo {author} {\bibfnamefont {C.}~\bibnamefont {Ciuti}}, \bibinfo {author} {\bibfnamefont {T.}~\bibnamefont {Coudreau}}, \bibinfo {author} {\bibfnamefont {P.}~\bibnamefont {Milman}}, \ and\ \bibinfo {author} {\bibfnamefont {S.}~\bibnamefont {Felicetti}},\ }\bibfield  {title} {\enquote {\bibinfo {title} {Superradiant phase transition in the ultrastrong-coupling regime of the two-photon {D}icke model},}\ }\href {\doibase 10.1103/PhysRevA.95.053854} {\bibfield  {journal} {\bibinfo  {journal} {Phys. Rev. A}\ }\textbf {\bibinfo {volume} {95}},\ \bibinfo {pages} {053854} (\bibinfo {year} {2017})}\BibitemShut {NoStop}%
\bibitem [{\citenamefont {Chen}\ and\ \citenamefont {Zhang}(2018)}]{Chen2018}%
  \BibitemOpen
  \bibfield  {author} {\bibinfo {author} {\bibfnamefont {Xiang-You}\ \bibnamefont {Chen}}\ and\ \bibinfo {author} {\bibfnamefont {Yu-Yu}\ \bibnamefont {Zhang}},\ }\bibfield  {title} {\enquote {\bibinfo {title} {Finite-size scaling analysis in the two-photon {D}icke model},}\ }\href {\doibase 10.1103/PhysRevA.97.053821} {\bibfield  {journal} {\bibinfo  {journal} {Phys. Rev. A}\ }\textbf {\bibinfo {volume} {97}},\ \bibinfo {pages} {053821} (\bibinfo {year} {2018})}\BibitemShut {NoStop}%
\bibitem [{\citenamefont {Garbe}\ \emph {et~al.}(2020)\citenamefont {Garbe}, \citenamefont {Wade}, \citenamefont {Minganti}, \citenamefont {Shammah}, \citenamefont {Felicetti},\ and\ \citenamefont {Nori}}]{Garbe2020}%
  \BibitemOpen
  \bibfield  {author} {\bibinfo {author} {\bibfnamefont {Louis}\ \bibnamefont {Garbe}}, \bibinfo {author} {\bibfnamefont {Peregrine}\ \bibnamefont {Wade}}, \bibinfo {author} {\bibfnamefont {Fabrizio}\ \bibnamefont {Minganti}}, \bibinfo {author} {\bibfnamefont {Nathan}\ \bibnamefont {Shammah}}, \bibinfo {author} {\bibfnamefont {Simone}\ \bibnamefont {Felicetti}}, \ and\ \bibinfo {author} {\bibfnamefont {Franco}\ \bibnamefont {Nori}},\ }\bibfield  {title} {\enquote {\bibinfo {title} {Dissipation-induced bistability in the two-photon {D}icke model},}\ }\href {\doibase 10.1038/s41598-020-69704-6} {\bibfield  {journal} {\bibinfo  {journal} {Sci. Rep.}\ }\textbf {\bibinfo {volume} {10}},\ \bibinfo {pages} {13408} (\bibinfo {year} {2020})}\BibitemShut {NoStop}%
\bibitem [{\citenamefont {Jayesh~Shah}\ \emph {et~al.}(2025)\citenamefont {Jayesh~Shah}, \citenamefont {Kirton}, \citenamefont {Felicetti},\ and\ \citenamefont {Alaeian}}]{Shah2025}%
  \BibitemOpen
  \bibfield  {author} {\bibinfo {author} {\bibfnamefont {Aanal}\ \bibnamefont {Jayesh~Shah}}, \bibinfo {author} {\bibfnamefont {Peter}\ \bibnamefont {Kirton}}, \bibinfo {author} {\bibfnamefont {Simone}\ \bibnamefont {Felicetti}}, \ and\ \bibinfo {author} {\bibfnamefont {Hadiseh}\ \bibnamefont {Alaeian}},\ }\bibfield  {title} {\enquote {\bibinfo {title} {Dissipative phase transition in the two-photon {D}icke model},}\ }\href {\doibase 10.1103/mz92-6l9g} {\bibfield  {journal} {\bibinfo  {journal} {Phys. Rev. Lett.}\ }\textbf {\bibinfo {volume} {135}},\ \bibinfo {pages} {173602} (\bibinfo {year} {2025})}\BibitemShut {NoStop}%
\bibitem [{\citenamefont {Zhai}\ \emph {et~al.}(2025)\citenamefont {Zhai}, \citenamefont {Wu}, \citenamefont {Wu},\ and\ \citenamefont {Chen}}]{Zhai2025}%
  \BibitemOpen
  \bibfield  {author} {\bibinfo {author} {\bibfnamefont {Cui-Lu}\ \bibnamefont {Zhai}}, \bibinfo {author} {\bibfnamefont {Wei}\ \bibnamefont {Wu}}, \bibinfo {author} {\bibfnamefont {Chun-Wang}\ \bibnamefont {Wu}}, \ and\ \bibinfo {author} {\bibfnamefont {Ping-Xing}\ \bibnamefont {Chen}},\ }\bibfield  {title} {\enquote {\bibinfo {title} {Stark-induced tunable phase transition in the two-photon {D}icke-{S}tark model},}\ }\href {\doibase 10.1103/vqnf-zd5w} {\bibfield  {journal} {\bibinfo  {journal} {Phys. Rev. A}\ }\textbf {\bibinfo {volume} {112}},\ \bibinfo {pages} {013720} (\bibinfo {year} {2025})}\BibitemShut {NoStop}%
\bibitem [{foo(See Supplemental Material at~\url{http://link.aps.org/supplemental/10.1103/syq9-z93n} for additional information about the two-photon Dicke model, which includes Refs.~\cite{DeAguiar1992,Bakemeier2013,Bastarrachea2014a,Bastarrachea2014b,Bastarrachea2015,Chavez2016})}]{footSM}%
  \BibitemOpen
  \href@noop {} {} (\bibinfo {year} {See Supplemental Material at~\url{http://link.aps.org/supplemental/10.1103/syq9-z93n} for additional information about the two-photon Dicke model, which includes Refs.~\cite{DeAguiar1992,Bakemeier2013,Bastarrachea2014a,Bastarrachea2014b,Bastarrachea2015,Chavez2016}})\BibitemShut {NoStop}%
\bibitem [{\citenamefont {{de Aguiar}}\ \emph {et~al.}(1992)\citenamefont {{de Aguiar}}, \citenamefont {Furuya}, \citenamefont {Lewenkopf},\ and\ \citenamefont {Nemes}}]{DeAguiar1992}%
  \BibitemOpen
  \bibfield  {author} {\bibinfo {author} {\bibfnamefont {M.A.M}\ \bibnamefont {{de Aguiar}}}, \bibinfo {author} {\bibfnamefont {K}~\bibnamefont {Furuya}}, \bibinfo {author} {\bibfnamefont {C.H}\ \bibnamefont {Lewenkopf}}, \ and\ \bibinfo {author} {\bibfnamefont {M.C}\ \bibnamefont {Nemes}},\ }\bibfield  {title} {\enquote {\bibinfo {title} {Chaos in a spin-boson system: Classical analysis},}\ }\href {\doibase https://doi.org/10.1016/0003-4916(92)90178-O} {\bibfield  {journal} {\bibinfo  {journal} {Annals of Physics}\ }\textbf {\bibinfo {volume} {216}},\ \bibinfo {pages} {291--312} (\bibinfo {year} {1992})}\BibitemShut {NoStop}%
\bibitem [{\citenamefont {Bakemeier}\ \emph {et~al.}(2013)\citenamefont {Bakemeier}, \citenamefont {Alvermann},\ and\ \citenamefont {Fehske}}]{Bakemeier2013}%
  \BibitemOpen
  \bibfield  {author} {\bibinfo {author} {\bibfnamefont {L.}~\bibnamefont {Bakemeier}}, \bibinfo {author} {\bibfnamefont {A.}~\bibnamefont {Alvermann}}, \ and\ \bibinfo {author} {\bibfnamefont {H.}~\bibnamefont {Fehske}},\ }\bibfield  {title} {\enquote {\bibinfo {title} {Dynamics of the {D}icke model close to the classical limit},}\ }\href {\doibase 10.1103/PhysRevA.88.043835} {\bibfield  {journal} {\bibinfo  {journal} {Phys. Rev. A}\ }\textbf {\bibinfo {volume} {88}},\ \bibinfo {pages} {043835} (\bibinfo {year} {2013})}\BibitemShut {NoStop}%
\bibitem [{\citenamefont {Bastarrachea-Magnani}\ \emph {et~al.}(2014{\natexlab{a}})\citenamefont {Bastarrachea-Magnani}, \citenamefont {Lerma-Hern\'andez},\ and\ \citenamefont {Hirsch}}]{Bastarrachea2014a}%
  \BibitemOpen
  \bibfield  {author} {\bibinfo {author} {\bibfnamefont {M.~A.}\ \bibnamefont {Bastarrachea-Magnani}}, \bibinfo {author} {\bibfnamefont {S.}~\bibnamefont {Lerma-Hern\'andez}}, \ and\ \bibinfo {author} {\bibfnamefont {J.~G.}\ \bibnamefont {Hirsch}},\ }\bibfield  {title} {\enquote {\bibinfo {title} {Comparative quantum and semiclassical analysis of atom-field systems. {I}. {D}ensity of states and excited-state quantum phase transitions},}\ }\href {\doibase 10.1103/PhysRevA.89.032101} {\bibfield  {journal} {\bibinfo  {journal} {Phys. Rev. A}\ }\textbf {\bibinfo {volume} {89}},\ \bibinfo {pages} {032101} (\bibinfo {year} {2014}{\natexlab{a}})}\BibitemShut {NoStop}%
\bibitem [{\citenamefont {Bastarrachea-Magnani}\ \emph {et~al.}(2014{\natexlab{b}})\citenamefont {Bastarrachea-Magnani}, \citenamefont {Lerma-Hern\'andez},\ and\ \citenamefont {Hirsch}}]{Bastarrachea2014b}%
  \BibitemOpen
  \bibfield  {author} {\bibinfo {author} {\bibfnamefont {M.~A.}\ \bibnamefont {Bastarrachea-Magnani}}, \bibinfo {author} {\bibfnamefont {S.}~\bibnamefont {Lerma-Hern\'andez}}, \ and\ \bibinfo {author} {\bibfnamefont {J.~G.}\ \bibnamefont {Hirsch}},\ }\bibfield  {title} {\enquote {\bibinfo {title} {Comparative quantum and semiclassical analysis of atom-field systems. {II}. {C}haos and regularity},}\ }\href {\doibase 10.1103/PhysRevA.89.032102} {\bibfield  {journal} {\bibinfo  {journal} {Phys. Rev. A}\ }\textbf {\bibinfo {volume} {89}},\ \bibinfo {pages} {032102} (\bibinfo {year} {2014}{\natexlab{b}})}\BibitemShut {NoStop}%
\bibitem [{\citenamefont {Bastarrachea-Magnani}\ \emph {et~al.}(2015)\citenamefont {Bastarrachea-Magnani}, \citenamefont {del Carpio}, \citenamefont {Lerma-Hern\'andez},\ and\ \citenamefont {Hirsch}}]{Bastarrachea2015}%
  \BibitemOpen
  \bibfield  {author} {\bibinfo {author} {\bibfnamefont {Miguel~Angel}\ \bibnamefont {Bastarrachea-Magnani}}, \bibinfo {author} {\bibfnamefont {Baldemar~L\'opez}\ \bibnamefont {del Carpio}}, \bibinfo {author} {\bibfnamefont {Sergio}\ \bibnamefont {Lerma-Hern\'andez}}, \ and\ \bibinfo {author} {\bibfnamefont {Jorge~G}\ \bibnamefont {Hirsch}},\ }\bibfield  {title} {\enquote {\bibinfo {title} {Chaos in the {D}icke model: Quantum and semiclassical analysis},}\ }\href {http://stacks.iop.org/1402-4896/90/i=6/a=068015} {\bibfield  {journal} {\bibinfo  {journal} {Phys. Scr.}\ }\textbf {\bibinfo {volume} {90}},\ \bibinfo {pages} {068015} (\bibinfo {year} {2015})}\BibitemShut {NoStop}%
\bibitem [{\citenamefont {Ch\'avez-Carlos}\ \emph {et~al.}(2016)\citenamefont {Ch\'avez-Carlos}, \citenamefont {Bastarrachea-Magnani}, \citenamefont {Lerma-Hern\'andez},\ and\ \citenamefont {Hirsch}}]{Chavez2016}%
  \BibitemOpen
  \bibfield  {author} {\bibinfo {author} {\bibfnamefont {J.}~\bibnamefont {Ch\'avez-Carlos}}, \bibinfo {author} {\bibfnamefont {M.~A.}\ \bibnamefont {Bastarrachea-Magnani}}, \bibinfo {author} {\bibfnamefont {S.}~\bibnamefont {Lerma-Hern\'andez}}, \ and\ \bibinfo {author} {\bibfnamefont {J.~G.}\ \bibnamefont {Hirsch}},\ }\bibfield  {title} {\enquote {\bibinfo {title} {Classical chaos in atom-field systems},}\ }\href {\doibase 10.1103/PhysRevE.94.022209} {\bibfield  {journal} {\bibinfo  {journal} {Phys. Rev. E}\ }\textbf {\bibinfo {volume} {94}},\ \bibinfo {pages} {022209} (\bibinfo {year} {2016})}\BibitemShut {NoStop}%
\end{thebibliography}%

\newpage
\cleardoublepage

\begin{widetext}

\setcounter{equation}{0}
\setcounter{figure}{0}

\makeatletter 
\renewcommand{\thesection}{S\@Roman\c@section}
\renewcommand{\thefigure}{S\@arabic\c@figure}
\renewcommand{\theequation}{S\@arabic\c@equation}
\makeatother

\begin{center}
    {\large \textbf{Supplemental Material: Superradiant Phase is a Finite Size Effect in Two-photon Processes} \\}
    \vspace{0.2in}
    Fabrizio Ram\'irez,$^{1}$ David Villase\~nor,$^{2}$ Nahum V\'azquez,$^{1}$ and Jorge G. Hirsch$^{1}$ \\
    \vspace{0.1in}
    {\small \it{$^{1}$Instituto de Ciencias Nucleares, Universidad Nacional Aut\'onoma de M\'exico, \\ Apdo. Postal 70-543, C.P. 04510 Mexico City, Mexico} \\}
    {\small \it{$^{2}$CAMTP - Center for Applied Mathematics and Theoretical Physics, \\ University of Maribor, Mladinska 3, SI-2000 Maribor, Slovenia} \\}
\end{center}

This Supplemental Material provides additional information on the two-photon Dicke model. In Sec.~\ref{sec:MeanFieldSqueezedStates}, we introduce the mean-field approximation using squeezed vacuum states, providing a classical limit for the system. The critical points of the equations of motion derived from this classical limit yield analytic expressions for the ground-state energy, the number of photons, and atomic inversion in both the normal and superradiant phases. In Sec.~\ref{sec:CollapseSuperradiantPhase}, we present further evidence regarding the disappearance of the superradiant phase, including the case of null atomic frequency. Section~\ref{sec:ThermodynamicLimitSqueezedStates} discusses the thermodynamic limit of the mean-field approximation with squeezed vacuum states, which helps us to detect the classical signature of the spectral collapse. Finally, in Sec.~\ref{sec:MeanFieldCoherentStates}, we perform a mean-field approximation using coherent states and describe the classical signature of the spectral collapse using this approach.

\section{Mean-field approximation with squeezed vacuum states}
\label{sec:MeanFieldSqueezedStates}

Inspired by the results of the ground state of the two-photon Dicke Hamiltonian presented in Ref.~\cite{Garbe2017}, we introduce a mean-field approximation to define a classical phase space using squeezed vacuum states $|\xi\rangle$ for the bosonic sector
\begin{equation}
    |\xi\rangle = e^{(\xi^{\ast}\hat{a}^{2}-\xi\hat{a}^{\dagger2})/2}|0\rangle ,
\end{equation}
and Bloch coherent states $|\beta\rangle$ for the atomic sector
\begin{equation}
    \label{eq:BlochState}
    |\beta\rangle = \frac{1}{(1+|\beta|^{2})^{j}}e^{\beta\hat{J}_{+}}|j,-j\rangle
\end{equation}
with complex parameter
\begin{equation}
    \label{eq:AtomicParameter}
    \beta = \frac{Q+iP}{\sqrt{4-Q^{2}-P^{2}}} ,
\end{equation}
where $(Q,P)$ are the atomic classical variables of phase space. In the above expressions, $|0\rangle$ is the bosonic vacuum state and $|j,-j\rangle$ is the atomic state with all atoms in the ground state. The squeezing parameter $\xi = r e^{i\theta}\in\mathbb{C}$ is a complex number with magnitude $r$ and phase $\theta$. It provides a connection with bosonic classical variables $(q,p)$ using the transformation $q=\sqrt{2/j}\sinh(r)\cos(\theta)$ and $p=\sqrt{2/j}\sinh(r)\sin(\theta)$
\begin{align}
    \langle\xi|\hat{a}|\xi\rangle = & \langle\xi|\hat{a}^{\dagger}|\xi\rangle = 0, \\
    \langle\xi|\hat{a}^{2}|\xi\rangle = & \langle\xi|\hat{a}^{\dagger2}|\xi\rangle^{\ast} = -e^{i\theta}\sinh(r)\cosh(r) = -\sqrt{\frac{j}{2}\left(1+\frac{j}{2}\left(q^{2}+p^{2}\right)\right)}(q+ip) , \\
    \langle\xi|\hat{a}^{\dagger}\hat{a}|\xi\rangle = & \sinh^{2}(r) = \frac{j}{2}\left(q^{2}+p^{2}\right) .
\end{align}

The expectation value of the quantum Hamiltonian $\hat{H}_{\text{D}}$ [Eq.~(1) in the main text] under squeezed vacuum states $|\mathbf{x}\rangle \equiv |\xi\rangle\otimes|\beta\rangle$, where $\mathbf{x}=(q,p;Q,P)$, provides a classical limit of the two-photon Dicke model
\begin{align}
    \label{eq:DickeSqueezedStates}
    h_{\text{D}}(\mathbf{x}) = & \frac{1}{j}\langle\mathbf{x}|\hat{H}_{\text{D}}|\mathbf{x}\rangle = \frac{1}{j}\langle\beta|\otimes\langle\xi|\hat{H}_{\text{D}}|\xi\rangle\otimes|\beta\rangle \nonumber \\
    = & \frac{1}{j}\left\{\omega \langle\xi|\hat{a}^{\dagger}\hat{a}|\xi\rangle - j\omega_{0} \frac{1-|\beta|^{2}}{1+|\beta|^{2}} + \gamma \left[ \langle\xi|\hat{a}^{\dagger2}|\xi\rangle + \langle\xi|\hat{a}^{2}|\xi\rangle \right] \frac{\beta+\beta^{\ast}}{1+|\beta|^{2}}\right\}\nonumber \\
    = & \frac{1}{j}\left\{\omega \sinh^{2}(r) + j\omega_{0}\left(\frac{Q^{2}+P^{2}}{2}-1\right)  - 2\gamma\sinh(r)\cosh(r)\cos(\theta) Q\sqrt{1-\frac{Q^{2}+P^{2}}{4}} \right\}  \nonumber 
    \\
    = & \frac{\omega}{2}\left(q^{2}+p^{2}\right) + \omega_{0}\left(\frac{Q^{2}+P^{2}}{2}-1\right) - 2\gamma qQ\sqrt{\left(\frac{1}{2j}+\frac{q^{2}+p^{2}}{4}\right)\left(1-\frac{Q^{2}+P^{2}}{4}\right)} ,
\end{align}
where the energy shells $\epsilon=E/j$ are scaled to the system size $j$.

The Hamilton's equations of motion of the Hamiltonian presented in Eq.~\eqref{eq:DickeSqueezedStates} are given by
\begin{align}
    \dot{q} = & \frac{\partial h_{\text{D}}(\mathbf{x})}{\partial p} = \omega p - \frac{\gamma qpQ}{2}\sqrt{1-\frac{Q^{2} + P^{2}}{4}}\left(\frac{1}{2j}+\frac{q^{2}+p^{2}}{4}\right)^{-1/2} , \\
    \dot{p} = & -\frac{\partial h_{\text{D}}(\mathbf{x})}{\partial q} = - \omega q + 2\gamma Q\sqrt{1-\frac{Q^{2} + P^{2}}{4}}\left(\frac{1}{2j}+\frac{2q^{2}+p^{2}}{4}\right)\left(\frac{1}{2j}+\frac{q^{2}+p^{2}}{4}\right)^{-1/2} , \\
    \dot{Q} = & \frac{\partial h_{\text{D}}(\mathbf{x})}{\partial P} = \omega_{0}P + \frac{\gamma qQP}{2}\sqrt{\frac{1}{2j}+\frac{q^{2}+p^{2}}{4}}\left(1-\frac{Q^{2} + P^{2}}{2}\right)^{-1/2} , \\
    \dot{P} = & -\frac{\partial h_{\text{D}}(\mathbf{x})}{\partial Q} = -\omega_{0}Q + 2\gamma q\sqrt{\frac{1}{2j}+\frac{q^{2}+p^{2}}{4}}\left(1-\frac{2Q^{2} + P^{2}}{4}\right)\left(1-\frac{Q^{2} + P^{2}}{4}\right)^{-1/2} . 
\end{align}
The minimization of the last set of equations $\dot{q}=\dot{p}=\dot{Q}=\dot{P}=0$ provides the critical points
\begin{align}
    \label{eq:ExtremeValuesNP}
    \mathbf{x}_{0} = & (0,0;0,0)  \hspace{0.5in} \text{for (NP)}, \\
    \label{eq:ExtremeValuesSPp}
    \mathbf{x}_{+} = & \left(\pm\frac{1}{\sqrt{j}}\left(\sqrt{\frac{1-\frac{4\gamma^{2}}{\omega^{2}}}{1-\frac{j^{2}\omega_{0}^{2}}{\gamma^{2}}}}-1\right)^{1/2},0;+\left(2-\frac{j\omega_{0}\omega}{\gamma^{2}}\sqrt{\frac{1-\frac{4\gamma^{2}}{\omega^{2}}}{1-\frac{j^{2}\omega_{0}^{2}}{\gamma^{2}}}}\right)^{1/2},0\right) \hspace{0.5in} \text{for (SP)}, \\
    \label{eq:ExtremeValuesSPm}
    \mathbf{x}_{-} = & \left(\pm\frac{1}{\sqrt{j}}\left(\sqrt{\frac{1-\frac{4\gamma^{2}}{\omega^{2}}}{1-\frac{j^{2}\omega_{0}^{2}}{\gamma^{2}}}}-1\right)^{1/2},0;-\left(2-\frac{j\omega_{0}\omega}{\gamma^{2}}\sqrt{\frac{1-\frac{4\gamma^{2}}{\omega^{2}}}{1-\frac{j^{2}\omega_{0}^{2}}{\gamma^{2}}}}\right)^{1/2},0\right) \hspace{0.5in} \text{for (SP)},
\end{align}
where the solution $\mathbf{x}_{0}\in\mathbb{R}$ belongs to the normal phase (NP) and the solutions $\mathbf{x}_{\pm}\in\mathbb{R}$ belong to the superradiant phase (SP) and explicitly depend on the system size $j$.

By evaluating the Hamiltonian of Eq.~\eqref{eq:DickeSqueezedStates} with the extreme values $ \mathbf{x}_{m}=\mathbf{x}_{0,\pm}$ [Eqs.~\eqref{eq:ExtremeValuesNP}-\eqref{eq:ExtremeValuesSPm}], we find the ground-state energy $E_{0}=j \, h_{\text{D}}(\mathbf{x}_{m})$ presented in Eq.~(5) in the main text. The last result agrees with the description of the Holstein-Primakoff transformation presented in Ref.~\cite{Garbe2017}. Moreover, the expectation values of operators $\hat{a}^{\dagger}\hat{a}$ and $\hat{J}_{z}$ under the squeezed vacuum states provide the expressions
\begin{align}
   \langle\mathbf{x}|\hat{a}^{\dagger}\hat{a}|\mathbf{x}\rangle & = \langle\beta|\otimes\langle\xi|\hat{a}^{\dagger}\hat{a}|\xi\rangle\otimes|\beta\rangle = \frac{j}{2}\left(q^{2}+p^{2}\right) , \\
   \frac{\langle\mathbf{x}|\hat{J}_{z}|\mathbf{x}\rangle}{j} & = \frac{1}{j}\langle\beta|\otimes\langle\xi|\hat{J}_{z}|\xi\rangle\otimes|\beta\rangle = \frac{Q^{2}+P^{2}}{2}-1 .
\end{align}
By evaluating the last expressions with the extreme values $\mathbf{x}_{m}$, we find the corresponding Eqs.~(6) and~(7) in the main text.

\begin{figure}[ht]
    \centering
    \includegraphics[width=1\columnwidth]{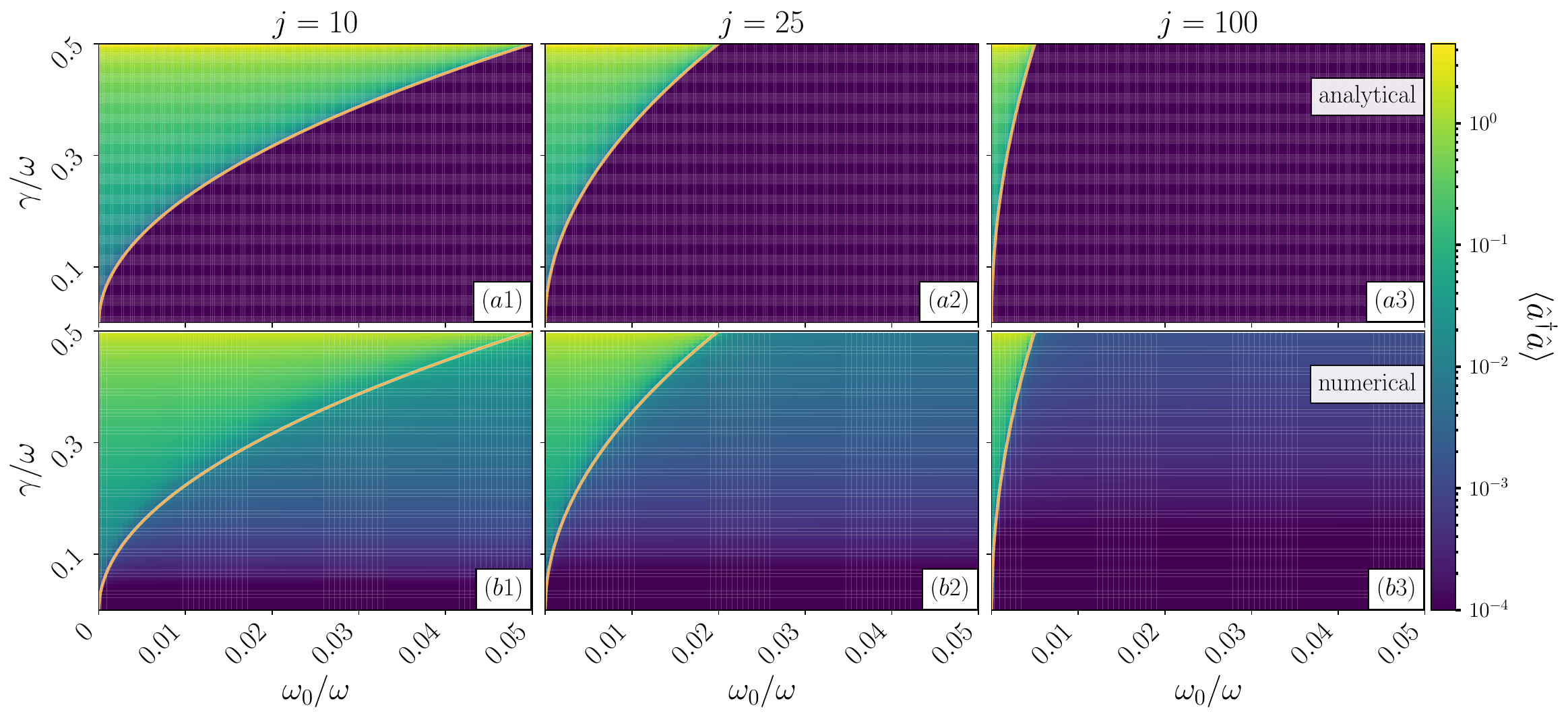}
    \caption{(a1)-(a3) Phase diagrams of the number of photons $\langle \hat{a}^{\dagger}\hat{a} \rangle$ [Eq.~(6) in the main text] as a function of the scaled atomic frequency $\omega_{0}/\omega$ and the scaled coupling strength $\gamma/\omega$. The color scheme represents the value of the number of photons in a logarithmic scale. The orange solid line represents the scaled critical coupling $\gamma_{\text{c}}/\omega=\sqrt{\omega_{0}j/(2\omega)}$. (b1)-(b3) Numerical implementations of panels (a1)-(a3). Each column identifies a different system size: (a1),(b1) $j=10$, (a2),(b2) $j=25$, and (a3),(b3) $j=100$.
    }
    \label{fig:nFadingPhase}
\end{figure}

\begin{figure}[ht]
    \centering
    \includegraphics[width=1\columnwidth]{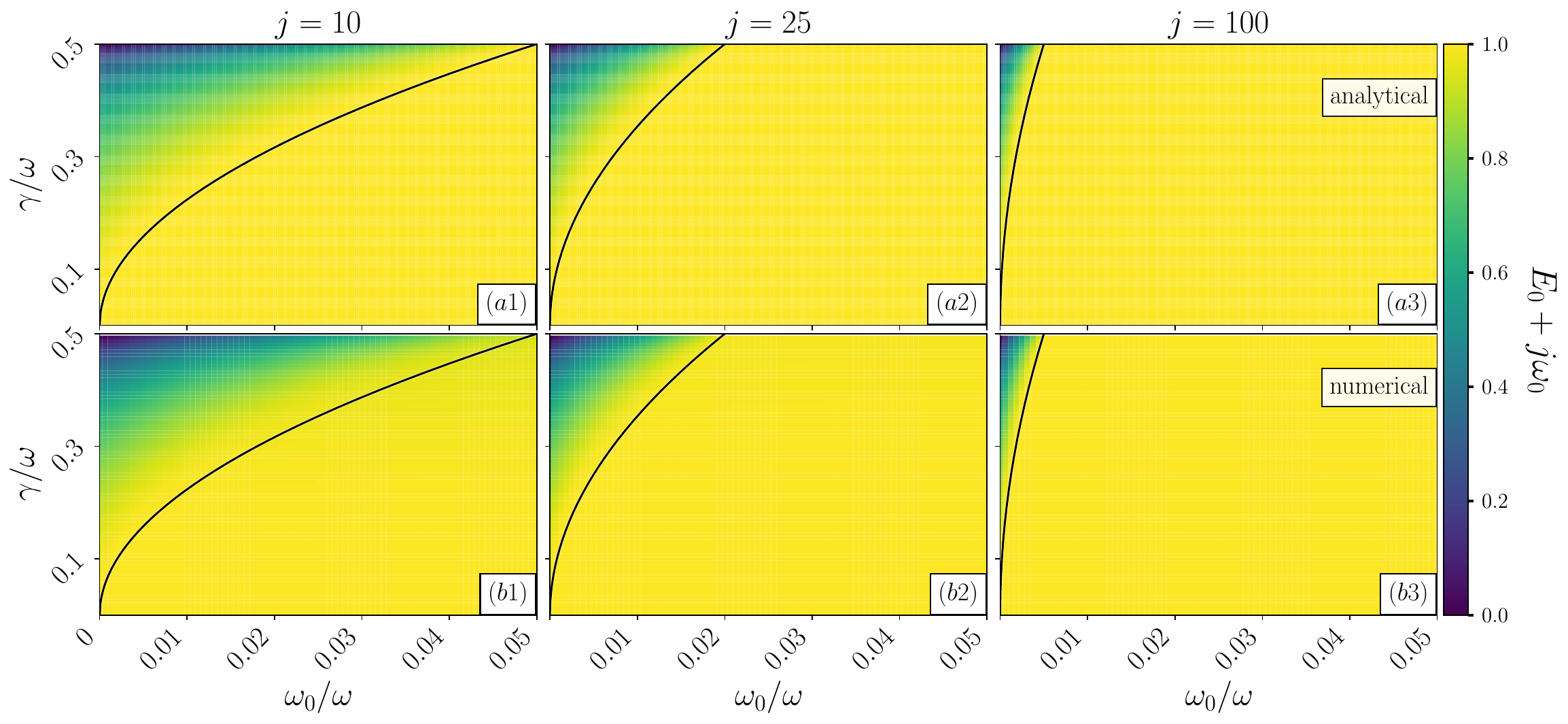}
    \caption{(a1)-(a3) Phase diagrams of the displaced ground-state energy $E_{0} + j|\omega_{0}|$ [Eq.~(5) in the main text] as a function of the scaled atomic frequency $\omega_{0}/\omega$ and the scaled coupling strength $\gamma/\omega$. The color scheme represents the value of the displaced ground-state energy. The black solid line represents the scaled critical coupling $\gamma_{\text{c}}/\omega=\sqrt{\omega_{0}j/(2\omega)}$. (b1)-(b3) Numerical implementations of panels (a1)-(a3). Each column identifies a different system size: (a1),(b1) $j=10$, (a2),(b2) $j=25$, and (a3),(b3) $j=100$.
    }
    \label{fig:EjFadingPhase}
\end{figure}

\begin{figure}[ht]
    \centering
    \includegraphics[width=1\columnwidth]{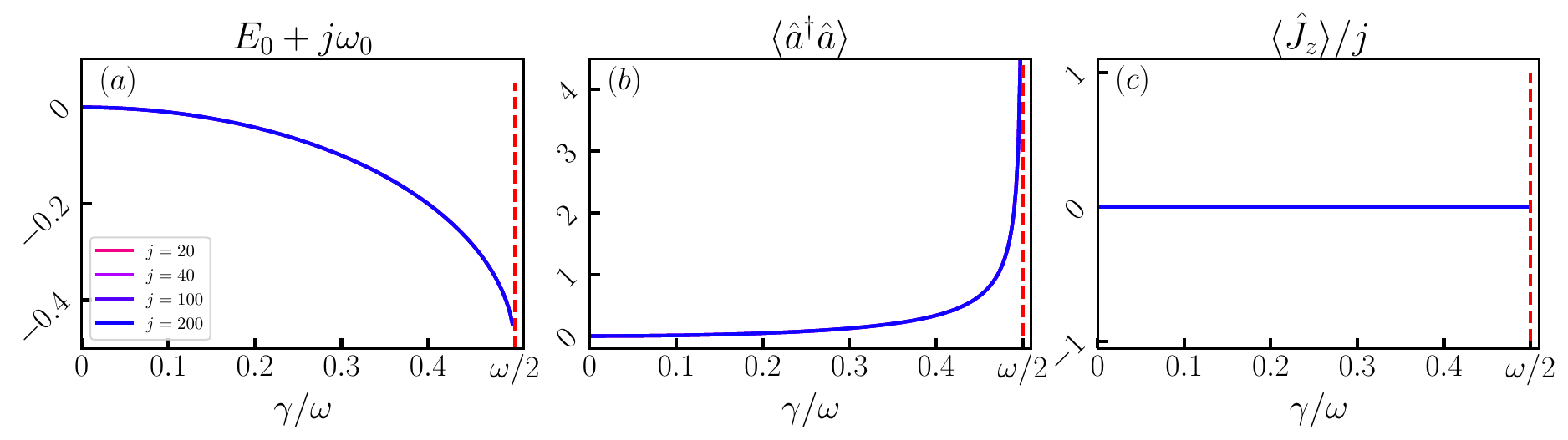}
    \caption{Projections of (a) the displaced ground-state energy $E_{0} + j|\omega_{0}|$ [Eq.~(5) in the main text], (b) the number of photons $\langle \hat{a}^{\dagger}\hat{a} \rangle$ [Eq.~(6) in the main text], and (c) the atomic inversion $\langle \hat{J}_{z} \rangle/j$ [Eq.~(7) in the main text] as a function of the scaled coupling strength $\gamma/\omega$. The vertical red dashed line represents the spectral collapse at $\gamma/\omega=0.5$. The atomic frequency is $\omega_{0}/\omega=0$.
    }
    \label{fig:PhaseDiagramSlices}
\end{figure}

\section{Disappearance of the superradiant phase and case with null atomic frequency}
\label{sec:CollapseSuperradiantPhase}

We show further evidence of the disappearance of the superradiant phase in the two-photon Dicke model. Similar to the phase diagrams presented in the main text for the normalized atomic excitation, we present phase diagrams for the number of photons and the ground-state energy.

In Fig.~\ref{fig:nFadingPhase}, we show phase diagrams of the number of photons $\langle\hat{a}^{\dagger}\hat{a}\rangle$ [Eq.~(6) in the main text] as a function of the scaled atomic frequency $\omega_{0}/\omega$ and the scaled coupling strength $\gamma/\omega$. The first row [Figs.~\ref{fig:nFadingPhase}(a1)-\ref{fig:nFadingPhase}(a3)] shows the number of photons for three values of the system size $j$ that increase from left to right. The color scheme represents the value of the number of photons in a logarithmic scale for the sake of clarity. The second row [Figs.~\ref{fig:nFadingPhase}(b1)-\ref{fig:nFadingPhase}(b3)] shows the numerical implementation using the ground state from the diagonalization of the Hamiltonian $\hat{H}_{\text{D}}$ [Eq.~(1) in the main text]. Here, we see how the numerical noise in the normal phase is more evident when compared with the analytical results [Figs.~\ref{fig:nFadingPhase}(a1)-\ref{fig:nFadingPhase}(a3)]. In general, we again identify how the superradiant phase shrinks with increasing system size $j$.

Similar phase diagrams are presented in Fig.~\ref{fig:EjFadingPhase} for the ground-state energy $E_{0}$ [Eq.~(5) in the main text]. The ground-state energy was displaced by adding the value of the ground-state energy of the normal phase, $E_{0}+j|\omega_{0}|$. The first row [Figs.~\ref{fig:EjFadingPhase}(a1)-\ref{fig:EjFadingPhase}(a3)] and the second row [Figs.~\ref{fig:EjFadingPhase}(b1)-\ref{fig:EjFadingPhase}(b3)] identify the analytical and numerical implementations, respectively. The color scheme represents the value of the displaced ground-state energy. The figures confirm again that the superradiant phase vanishes by increasing the system size $j$.

When $\omega_{0}=0$, all the observables presented in Eqs.~(5)-(7) of the main text become independent of the system size $j$. In Fig.~\ref{fig:PhaseDiagramSlices}, we display these observables as a function of the scaled coupling strength $\gamma/\omega$. We illustrate the ground-state energy in Fig.~\ref{fig:PhaseDiagramSlices}(a), which has been displaced by the ground-state energy of the normal phase $E_{0}+ j|\omega_{0}|$. In Fig.~\ref{fig:PhaseDiagramSlices}(b), we present the number of photons $\langle\hat{a}^{\dagger}\hat{a}\rangle$,  Finally, in Fig.~\ref{fig:PhaseDiagramSlices}(c), we show the atomic inversion $\langle\hat{J}_{z}\rangle/j$. The red dashed line indicates the spectral collapse at $\gamma/\omega=0.5$ in all panels. In Fig.~\ref{fig:PhaseDiagramSlices}(a), we observe a finite value of the displaced ground-state energy when the spectral collapse occurs. Figure~\ref{fig:PhaseDiagramSlices}(b) shows a divergence in the number of photons at the point of spectral collapse, while in Fig.~\ref{fig:PhaseDiagramSlices}(c), we find that the atomic inversion has a constant mean value of zero, as expected. Notice that in this case the atomic degree of freedom becomes decoupled from the photonic one. The divergence in the photon number is a result of spectral collapse; it is independent of the number of photons and is not related to a superradiant phase.

\section{Thermodynamic limit with squeezed vacuum states and spectral collapse}
\label{sec:ThermodynamicLimitSqueezedStates}

Once we detected the disappearance of the superradiant phase in the thermodynamic limit, we focus now on the description of the spectral collapse in the same limit. The thermodynamic limit of Eq.~\eqref{eq:DickeSqueezedStates} provides the classical Hamiltonian
\begin{align}
    \label{eq:DickeSqueezedStatesInfinity}
    h_{\text{D}}^{\infty}(\mathbf{x}) = & \lim_{j\to\infty}h_{\text{D}}(\mathbf{x}) \nonumber \\
    = & \frac{\omega}{2}\left(q^{2}+p^{2}\right) + \omega_{0}\left(\frac{Q^{2}+P^{2}}{2}-1\right) - \gamma qQ\sqrt{\left(q^{2}+p^{2}\right)\left(1-\frac{Q^{2}+P^{2}}{4}\right)}  
\end{align}
and the Hamilton's equations of motion
\begin{align}
    \dot{q} = & \frac{\partial h_{\text{D}}^{\infty}(\mathbf{x})}{\partial p} = \omega p - \frac{\gamma qpQ}{\sqrt{q^{2}+p^{2}}}\sqrt{1-\frac{Q^{2} + P^{2}}{4}} , \\
    \dot{p} = & -\frac{\partial h_{\text{D}}^{\infty}(\mathbf{x})}{\partial q} = - \omega q + \frac{\gamma \left(2q^{2}+p^{2}\right)Q}{\sqrt{q^{2}+p^{2}}}\sqrt{1-\frac{Q^{2} + P^{2}}{4}} , \\
    \dot{Q} = & \frac{\partial h_{\text{D}}^{\infty}(\mathbf{x})}{\partial P} = \omega_{0}P + \frac{\gamma q\sqrt{q^{2}+p^{2}}QP}{4}\left(1-\frac{Q^{2} + P^{2}}{2}\right)^{-1/2} , \\
    \dot{P} = & -\frac{\partial h_{\text{D}}^{\infty}(\mathbf{x})}{\partial Q} = -\omega_{0}Q + \gamma q\sqrt{q^{2}+p^{2}}\left(1-\frac{2Q^{2} + P^{2}}{4}\right)\left(1-\frac{Q^{2} + P^{2}}{4}\right)^{-1/2} . 
\end{align}
The minimization of the last set of equations $\dot{q}=\dot{p}=\dot{Q}=\dot{P}=0$ gives the critical points
\begin{align}
    \label{eq:ExtremeValuesNPInfinity}
    \mathbf{x}_{0} = & (0,0;0,0)  \hspace{0.5in} \text{for (NP)}, \\
    \label{eq:ExtremeValuesSPpInfinity}
    \mathbf{x}_{+}^{\infty} = & \left(\pm(-1)^{1/4}\sqrt{\frac{\omega_{0}}{\gamma}}\left(1-\frac{4\gamma^{2}}{\omega^{2}}\right)^{-1/4},0;+\left(2+i\frac{\omega}{\gamma}\sqrt{1-\frac{4\gamma^{2}}{\omega^{2}}}\right)^{1/2},0\right) \hspace{0.5in} \text{for (SP)}, \\
    \label{eq:ExtremeValuesSPmInfinity}
    \mathbf{x}_{-}^{\infty} = & \left(\pm(-1)^{1/4}\sqrt{\frac{\omega_{0}}{\gamma}}\left(1-\frac{4\gamma^{2}}{\omega^{2}}\right)^{-1/4},0;-\left(2+i\frac{\omega}{\gamma}\sqrt{1-\frac{4\gamma^{2}}{\omega^{2}}}\right)^{1/2},0\right) \hspace{0.5in} \text{for (SP)},
\end{align}
where $\mathbf{x}_{0}\in\mathbb{R}$ and $\mathbf{x}_{\pm}^{\infty}\in\mathbb{C}$. The real point $\mathbf{x}_{0}$ is the only allowed solution that reproduces the ground-state energy of the normal phase, when we evaluate $h_{\text{D}}^{\infty}(\mathbf{x}_{0})=E_{0}/j$. The complex solutions $\mathbf{x}_{\pm}^{\infty}\in\mathbb{C}$ are unphysical and as a result the superradiant phase is not allowed. However, the Hamiltonian $h_{\text{D}}^{\infty}(\mathbf{x})$ can describe the classical signature of spectral collapse in the bosonic phase space, using the energy surface $h_{\text{D}}^{\infty}(q,p)$ [Eq.~(9) in the main text].

On the other hand, if we consider the critical points of the Hamiltonian $h_{\text{D}}(\mathbf{x})$ [Eqs.~\eqref{eq:ExtremeValuesNP}-\eqref{eq:ExtremeValuesSPm}] and take the thermodynamic limit $j\to\infty$, we arrive to the same critical points of the Hamiltonian $h_{\text{D}}^{\infty}(\mathbf{x})$ [Eqs.~\eqref{eq:ExtremeValuesNPInfinity}-\eqref{eq:ExtremeValuesSPmInfinity}]
\begin{align}
    \lim_{j\to\infty}\mathbf{x}_{0} = \mathbf{x}_{0}\in\mathbb{R} \hspace{0.5in} \text{and} \hspace{0.5in} \lim_{j\to\infty}\mathbf{x}_{\pm} = \mathbf{x}_{\pm}^{\infty}\in\mathbb{C},
\end{align}
signaling how the superradiant-phase solutions $\mathbf{x}_{\pm}\in\mathbb{R}$ become complex $\mathbf{x}_{\pm}^{\infty}\in\mathbb{C}$ in the thermodynamic limit. This also can be seen with the analysis of the thermodynamic limit of the ground-state energy, the number of photons, and the atomic inversion [Eqs.~(5)-(7) in the main text], all of them properly scaled by the system size,
\begin{align}
     \lim_{j\to\infty}\frac{E_{0}}{j} = \lim_{j\to\infty}\frac{\langle \mathbf{x}_{m}| \hat{H}_{\text{D}} |\mathbf{x}_{m} \rangle}{j} & = -\left\{\begin{array}{ll}
       |\omega_{0}| & \text{for (NP)} \\
        i\frac{\omega\omega_{0}}{2\gamma}\sqrt{1-\frac{4\gamma^{2}}{\omega^{2}}} & \text{for (SP)}
    \end{array}\right. , \\
    \lim_{j\to\infty}\frac{\langle \mathbf{x}_{m}| \hat{a}^{\dagger}\hat{a} |\mathbf{x}_{m} \rangle}{j} & = \left\{\begin{array}{ll}
        0 & \text{for (NP)} \\
        -i\frac{\omega_{0}}{2\gamma}\left(1-\frac{4\gamma^{2}}{\omega^{2}}\right)^{-1/2} & \text{for (SP)}
    \end{array}\right. , \\
    \lim_{j\to\infty}\frac{\langle \mathbf{x}_{m}| \hat{J}_{z} |\mathbf{x}_{m} \rangle}{j} & = -\left\{\begin{array}{ll}
        \text{sgn}(\omega_{0}) & \text{for (NP)} \\
        -i\frac{\omega}{2\gamma}\sqrt{1-\frac{4\gamma^{2}}{\omega^{2}}} & 
        \text{for (SP)}
    \end{array}\right. .
\end{align}
In the above expressions, we can see how the thermodynamic limit in the normal phase is well defined, while the thermodynamic limit in the superradiant phase gives complex expressions for all observables, providing further evidence about the impossibility of this phase.

\section{Mean-field approximation with coherent states}
\label{sec:MeanFieldCoherentStates}

Usual mean-field approximations for spin-boson systems consider coherent states as trial states~\cite{DeAguiar1992,Bakemeier2013,Bastarrachea2014a,Bastarrachea2014b,Bastarrachea2015,Chavez2016}. We consider Glauber coherent states $|\alpha\rangle$ to describe the bosonic sector
\begin{equation}
    |\alpha\rangle = e^{-|\alpha|^{2}/2}e^{\alpha\hat{a}^{\dagger}}|0\rangle ,
\end{equation}
with complex parameter
\begin{equation}
    \alpha = \sqrt{\frac{j}{2}}(q+ip) ,
\end{equation}
and Bloch coherent states $|\beta\rangle$ to describe the atomic sector [Eqs.~\eqref{eq:BlochState} and~\eqref{eq:AtomicParameter}].

The expectation value of the quantum Hamiltonian $\hat{H}_{\text{D}}$ [Eq.~(1) in the main text] under Glauber-Bloch coherent states $|\mathbf{y}\rangle \equiv |\alpha\rangle\otimes|\beta\rangle$, where $\mathbf{y}=(q,p;Q,P)$, provides a classical limit of the two-photon Dicke model~\cite{Ramirez2025}
\begin{align}
    \label{eq:DickeCoherentStates}
    h_{\text{D}}(\mathbf{y}) = & \frac{1}{j}\langle\mathbf{y}|\hat{H}_{\text{D}}|\mathbf{y}\rangle = \frac{1}{j}\langle\beta|\otimes\langle\alpha|\hat{H}_{\text{D}}|\alpha\rangle\otimes|\beta\rangle \nonumber \\
    = & \frac{1}{j}\left\{ \omega|\alpha|^{2}-j\omega_{0}\frac{1-|\beta|^{2}}{1+|\beta|^{2}}+\gamma\left[(\alpha^{\ast})^{2}+\alpha^{2}\right]\frac{\beta+\beta^{\ast}}{1+|\beta|^{2}} \right\} \nonumber \\
    = & \frac{\omega}{2}\left(q^{2}+p^{2}\right) + \omega_{0}\left(\frac{Q^{2}+P^{2}}{2}-1\right) + \gamma\left(q^{2}-p^{2}\right)Q\sqrt{1-\frac{Q^{2}+P^{2}}{4}} .
\end{align}

The Hamilton's equations of motion of the Hamiltonian presented in Eq.~\eqref{eq:DickeCoherentStates} are described by
\begin{align}
    \dot{q} = & \frac{\partial h_{\text{D}}(\mathbf{y})}{\partial p} = \omega p - 2\gamma pQ\sqrt{1-\frac{Q^{2} + P^{2}}{4}} , \\
    \dot{p} = & -\frac{\partial h_{\text{D}}(\mathbf{y})}{\partial q} = - \omega q - 2\gamma qQ\sqrt{1-\frac{Q^{2} + P^{2}}{4}} , \\
    \dot{Q} = & \frac{\partial h_{\text{D}}(\mathbf{y})}{\partial P} = \omega_{0}P - \frac{\gamma\left(q^{2}-p^{2}\right)QP}{4}\left(1-\frac{Q^{2} + P^{2}}{4}\right)^{-1/2} , \\
    \dot{P} = & -\frac{\partial h_{\text{D}}(\mathbf{y})}{\partial Q} = -\omega_{0}Q - \gamma\left(q^{2}-p^{2}\right)\left(1-\frac{2Q^{2} + P^{2}}{4}\right)\left(1-\frac{Q^{2} + P^{2}}{4}\right)^{-1/2} .
\end{align}
The minimization of the last set of equations $\dot{q}=\dot{p}=\dot{Q}=\dot{P}=0$ provides the critical points
\begin{align}
    \label{eq:ComplexSolution0}
    \mathbf{y}_{0} = & \mathbf{x}_{0} \hspace{0.5in} \text{for (NP)}, \\
    \label{eq:ComplexSolutionP}
    \mathbf{y}_{\pm} = & \mathbf{x}_{\pm}^{\infty} \hspace{0.5in} \text{for (SP)},
\end{align}
which are given in Eqs.~\eqref{eq:ExtremeValuesNPInfinity}-\eqref{eq:ExtremeValuesSPmInfinity}.

The Hamiltonian $h_{\text{D}}(\mathbf{y})$ obtained with coherent states has been proven to be useful for describing the classical dynamics of the two-photon Dicke model~\cite{Ramirez2025}. However, we explore the classical signature of the spectral collapse in direct analogy with the study of the Hamiltonian $h_{\text{D}}^{\infty}(\mathbf{x})$ [Eq.~(8) in the main text] obtained with squeezed vacuum states. An energy surface projected on the bosonic variables can be obtained for the Hamiltonian $h_{\text{D}}(\mathbf{y})$, similar to Eq.~(9) in the main text. We eliminate the atomic variables by solving $\partial_{Q}h_{\text{D}}(\mathbf{y})=\partial_{P}h_{\text{D}}(\mathbf{y})=0$, whose solutions $Q_{m}=Q_{m}(q,p)$ and $P_{m}=0$ provide the energy surface
\begin{align}
\label{eq:EnergySurfaceCoherent}
    h_{\text{D}}(q,p) = & h_{\text{D}}(q,p;Q_{m},0) \nonumber \\
    = & \frac{\omega}{2}\left(q^{2}+p^{2}\right)-\omega_{0}\sqrt{1+\frac{\gamma^{2}}{\omega_{0}^{2}}\left(q^{2}-p^{2}\right)^{2}} .
\end{align}

In Figs.~\ref{fig:SpectralCollapseCoherent}(a1)-\ref{fig:SpectralCollapseCoherent}(a4), we show the energy surface for increasing values of the scaled coupling strength $\gamma/\omega$ at fixed resonance $\omega_{0}=2\omega$. Figures~\ref{fig:SpectralCollapseCoherent}(b1)-\ref{fig:SpectralCollapseCoherent}(b4) show the orthogonal projection of the energy surfaces over the bosonic plane $q-p$. For weak coupling, the surface is bounded and displays a single global minimum associated to the normal phase. For increasing values of $\gamma/\omega$ near the collapse value $\gamma_{\text{sc}}/\omega=1/2$, the surface becomes unbounded and develops extended regions at the ground-state energy, this time across both $q$ and $p$ axis [Figs.~\ref{fig:SpectralCollapseCoherent}(a4) and~\ref{fig:SpectralCollapseCoherent}(b4)]. The last is an unexpected feature, since the spectral collapse detected with the Hamiltonian $h_{\text{D}}^{\infty}(\mathbf{x})$ became unbounded across the $p$ axis only [Figs. 3(a4) and 3(b4) in the main text]. Despite the last discrepancy, both classical limits $h_{\text{D}}^{\infty}(\mathbf{x})$ and $h_{\text{D}}(\mathbf{y})$ show a transition from a bounded to an unbounded energy region, which is interpreted as a clear classical signature of spectral collapse in bosonic phase space.

\begin{figure*}[ht]
    \centering
    \includegraphics[width=0.9\textwidth]{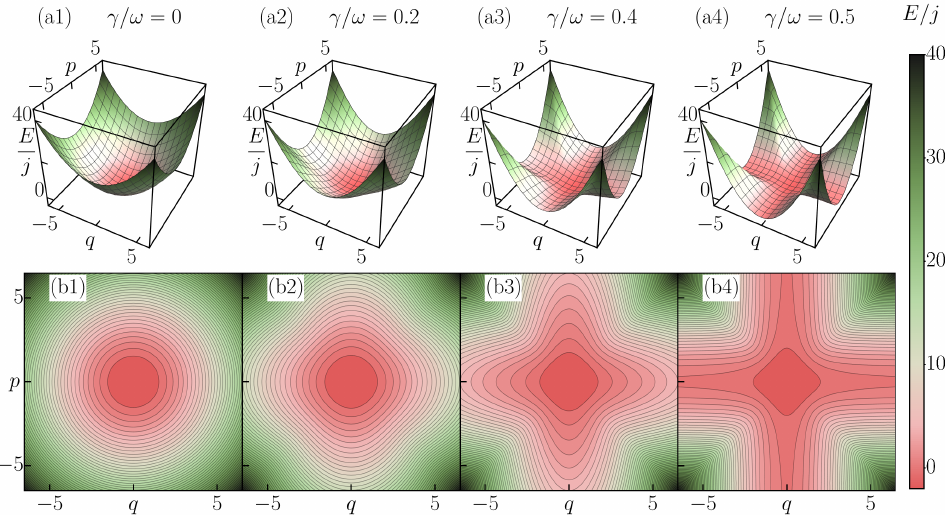}
    \caption{(a1)-(a4) Energy surface $h_{\text{D}}(q,p)$ [Eq.~\eqref{eq:EnergySurfaceCoherent}] as a function of the bosonic coordinates $(q,p)$ for increasing values of the scaled coupling strength $\gamma/\omega=0.1,0.2,0.4,0.5$. (b1)-(b4) Orthogonal projections of panels (a1)-(a4) over the bosonic plane $q-p$. The color scale shows values of the scaled classical energy $E/j$ that begins at the scaled ground-state energy $E_{0}/j=-\omega_{0}$. System parameters: $\omega=1$ and $\omega_{0}=2\omega$.}
    \label{fig:SpectralCollapseCoherent}
\end{figure*}

\end{widetext}

\end{document}